\documentclass[fleqn,11pt]{wlscirep}
\usepackage[utf8]{inputenc}
\usepackage[T1]{fontenc}

\usepackage{graphicx}
\usepackage{bm}
\usepackage{siunitx}
\usepackage{physics}
\usepackage{color}
\usepackage[skip=5pt, indent = 15pt]{parskip}

\usepackage[font=footnotesize]{caption}
\usepackage[numbers, super, sort&compress]{natbib}
\usepackage{bibunits}
\defaultbibliographystyle{naturemag-doi}
\defaultbibliography{main}

\usepackage{xurl}
\usepackage{hyperref}
\usepackage{lineno}

\def\thesection{\arabic{section}}
\def\thesubsection{\arabic{section}.\arabic{subsection}}
\makeatletter
\def\p@subsection{}
\def\p@subsubsection{}
\makeatother
\def\thesubsubsection{\arabic{section}.\arabic{subsection}.\arabic{subsubsection}}

\begin{document}
\title{Decoding cell signaling via optimal transport and information theory}

\author[1]{Mintu Nandi\thanks{mintunandi@ubi.s.u-tokyo.ac.jp}}
\author[1,2]{Sosuke Ito\thanks{sosuke.ito@ubi.s.u-tokyo.ac.jp}}
\affil[1]{Universal Biology Institute, The University of Tokyo, 7-3-1 Hongo, Bunkyo-ku, Tokyo 113-0033, Japan}
\affil[2]{Department of Physics, The University of Tokyo, 7-3-1 Hongo, Bunkyo-ku, Tokyo 113-0033, Japan}


\begin{abstract}

Cellular signal processing performs reliably despite molecular noise. Mutual information (MI) is widely used to quantify signaling fidelity, capturing how well outputs discriminate input states. However, it fails to capture whether the output preserves the statistical structure of the input, a property crucial in morphogen patterning and dose-dependent signaling. To address this gap, we introduce the 2-Wasserstein (2-WD) distance, which provides a geometric basis for comparing input and output distributions. We define MI as informational fidelity (INF) and the inverse of the 2-WD as geometric fidelity (GMF). Applying this dual-fidelity framework to canonical regulatory motifs under Gaussian channel approximation reveals topology-dependent trade-offs: coherent feed-forward loops can perform well in both dimensions, whereas feedback architectures reduce INF to enhance GMF. Experimental analysis of tumor necrosis factor signaling reveals dual-fidelity behavior qualitatively consistent with feedback regulation. RAS-MAPK data analysis further shows that jointly considering INF and GMF better characterizes intracellular signal relay than INF alone. Our results thus indicate that these signaling behaviors are not fully characterized by MI alone; instead, distributional correspondence provides a complementary dimension of signaling fidelity. Our study provides a practical framework for analyzing natural networks and guiding the design of task-specific synthetic circuits.
    
\end{abstract}

\maketitle


\begin{bibunit}

\section*{Introduction}
\label{sec1}

Cells sense, process, and respond to external and internal cues despite molecular noise. A widely adopted approach to quantify the reliability of signal transmission under noisy conditions is mutual information (MI) \cite{Shannon1948, Shannon1963, Waltermann2011, Uda2013}. An MI of 1 bit implies that the output can reliably distinguish between two equally likely input states \cite{Cheong2011, Uda2013, Tkacik2016}. As MI increases, the system gains higher resolution in distinguishing fine-grained input states, thus supporting more nuanced and precise cellular decisions \cite{Tostevin2009, Bowsher2014}.

\begin{figure*}[!t]
\centering
\includegraphics[width=0.95\columnwidth,angle=0]{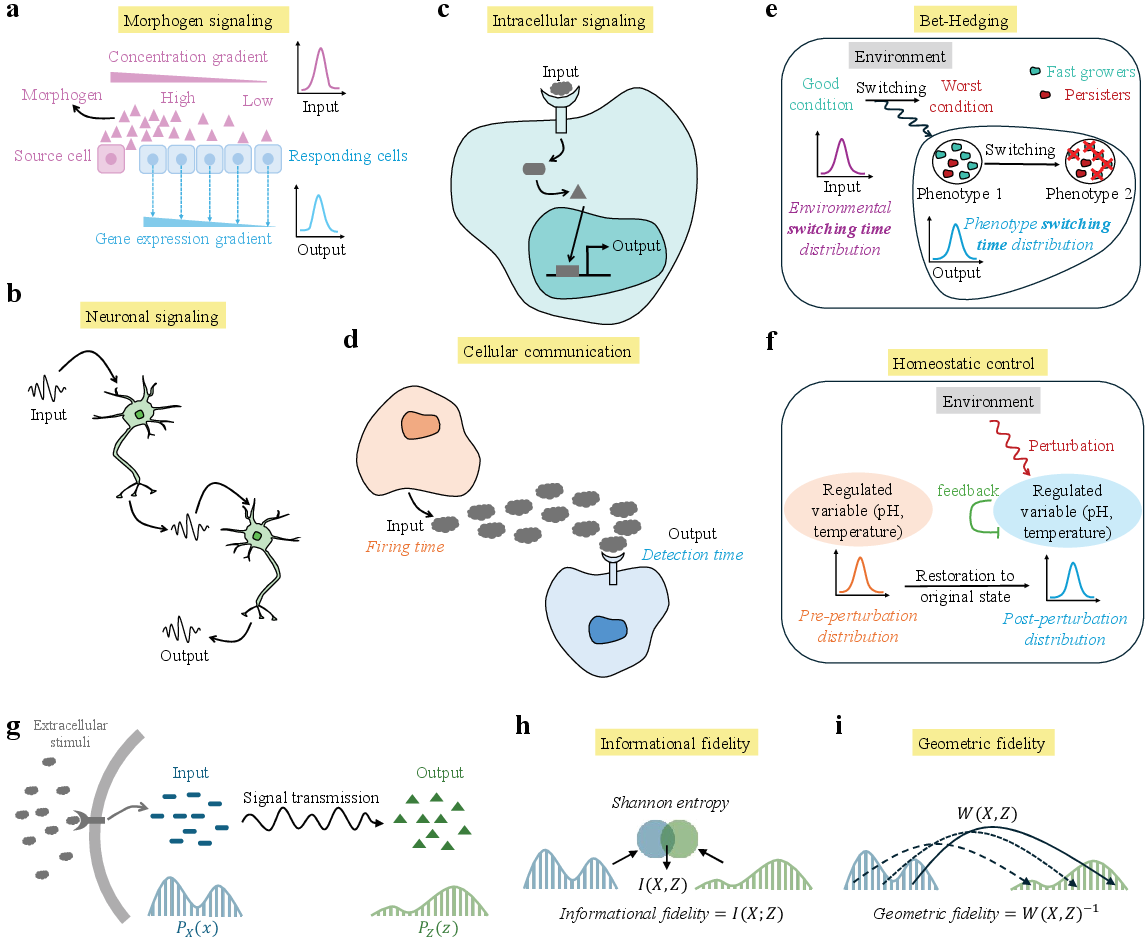}
\caption{\textbf{Schematics of signaling systems and the fidelity representations}.
Illustrative examples of signaling pathways: \textbf{a} morphogen signaling \cite{Gregor2007},
\textbf{b} neuronal signaling \cite{Lansky2023},
\textbf{c} intracellular signaling \cite{Takahashi2008, Kramar2025},
\textbf{d} cell-to-cell communication \cite{Sarkar2023},
\textbf{e} bet-hedging \cite{Kussell2005}, and
\textbf{f} homeostatic control \cite{Billman2020, Valls2022}.
\textbf{g} A general scheme of signal transduction. Here, the input $X$ acts as an internal representation of the external signals. The input distribution $P_X(x)$ is processed into an output $Z$ with distribution $P_Z(z)$.
\textbf{h} Informational fidelity is defined as MI between $X$ and $Z$, $I(X;Z)$. The MI is shown in terms of the overlap of Shannon entropies.
\textbf{i} Geometric fidelity is defined as the inverse of the 2-WD, $W(X,Z)^{-1}$. The 2-WD $W(X,Z)$ is introduced via an optimal transport problem.
}
\label{f1}
\end{figure*}

However, effective biological functions often require more than accurate state discrimination. The statistical alignment between input and output distributions can constitute a distinct dimension of performance. Such distributional correspondence is evident in developmental patterning, where morphogen-response profiles must be reproducible despite noise (Fig.~\ref{f1}a) \cite{Gregor2007}, and in dose-dependent signaling \cite{Takahashi2008, Kramar2025}, where cellular responses scale with input magnitude. Similarly, several signaling pathways evolve to avoid losing signaling features by transmitting the signal in a linear, undistorted manner \cite{Nunns2018, Andrews2018}. 

The biological meaning of distributional correspondence depends on the variables used to represent the input and output. In signal transduction, these variables are typically the concentrations or activities of signaling molecules and their downstream responses. In other contexts, they may represent timing in cellular communication \cite{Sarkar2023} (Fig.~\ref{f1}d), phenotypic states in bet-hedging \cite{Kussell2005} (Fig.~\ref{f1}e), or recovery distributions after perturbation in homeostasis (Fig.~\ref{f1}f) \cite{Billman2020, Valls2022}. Thus, distributional correspondence is task-dependent, but in all cases it captures whether the output preserves functionally relevant statistical features of the input. Here, we focus on signal transduction, where inputs and outputs correspond to signal and response concentrations.

While MI effectively captures how well input states can be distinguished by output, it fails to quantify input-output distributional correspondence. This limitation arises because MI is invariant under invertible transformations of the input or output \cite{Cover1991}. Consequently, MI can identify that a signal was received but cannot determine whether the output distribution preserves the quantitative structure of the input. These observations motivate treating input-output distributional correspondence as a distinct and biologically meaningful dimension of signaling fidelity.

To quantify the input-output correspondence, we introduce the 2-Wasserstein distance (2-WD) from optimal transport (OT) theory \cite{Villani2008}. It quantifies the minimum cost of transforming one probability distribution into another, offering a principled, geometric approach to comparing input and output distributions. In biochemical signaling, where the output is a regulated transformation of the input, the 2-WD value naturally captures how faithfully the input distribution is preserved downstream. A small 2-WD indicates that the output retains key quantitative features of the input, such as its shape, scale (variability), and mean. Although OT has been applied to developmental trajectories \cite{Schiebinger2019}, multi-omic alignment \cite{Demetci2022}, single-cell integration \cite{Cao2022}, and neuroscience \cite{Kawakita2024}, it remains underutilized in biochemical signal transduction.

Building on these perspectives, we integrate OT with information theory to capture the complementary aspects of signal transmission: state resolution and distributional correspondence. We hypothesize that optimal signal transduction can be achieved when these two aspects are tuned according to the demands of cellular functions. We consider a general input ($X$)--output ($Z$) signaling channel (Fig.~\ref{f1}g) and formulate a theoretical framework where MI, $I(X;Z)$, quantifies informational fidelity (INF)---how efficiently the output ($Z$) encodes the input states ($X$) (Fig.~\ref{f1}h). Moreover, the inverse of the 2-WD between the input and output distributions, $W(X,Z)^{-1}$, captures geometric fidelity (GMF)---how closely the output mirrors the input distribution (Fig.~\ref{f1}i). To apply this framework, we developed gene regulation models that allow for the analytical calculation of MI and 2-WD across canonical network motifs. This enables us to analyze how regulatory topology shapes cell signaling in the dual-fidelity landscape.

Here, we show that distinct network topologies navigate INF and GMF in motif-specific ways, revealing diverse signal processing strategies embedded in gene regulatory architectures. Notably, our analysis of tumor necrosis factor (TNF) signaling data reveals dual-fidelity shifts that are qualitatively consistent with the motif-specific trends predicted by the model. Analysis of RAS-MAPK signaling further shows that intracellular signal relay is more completely characterized by considering both INF and GMF than by INF alone. Together, these findings indicate that distributional correspondence provides an additional dimension of signaling behavior captured by GMF, beyond the input-output correlations reflected in INF. Thus, by extending conventional approaches that rely solely on MI, our framework provides a practical method for studying natural networks and guiding the design of synthetic circuits.

\section*{Results}
\label{sec2}

\noindent \textbf{The dual-fidelity framework.} To formalize the framework, we define a variational objective function that integrates MI, $I(X;Z)$, and the 2-WD, $W(X,Z)$, between the input $X$ and the output $Z$, expressed by a Lagrangian of the form:
\begin{equation}
    \mathcal{L} = I(X;Z) - \lambda [W(X,Z)]^2,
    \label{eqm1}
\end{equation}

\noindent where $\lambda$ is a Lagrange multiplier, with units of bits/(unit of $X$ or $Z$)$^2$. A low $\lambda$ weights MI more heavily, whereas a high $\lambda$ prioritizes the distributional correspondence between input and output. Note that this Lagrangian $\mathcal{L}$ is similar in spirit to the Sinkhorn distance formulation~\cite{cuturi2013sinkhorn}, which involves a trade-off between MI and optimal transport cost~\cite{feydy2019interpolating}. There is, however, a subtle distinction: the Sinkhorn formulation introduces this trade-off before optimization of transport cost, whereas the present Lagrangian considers it after optimization of transport cost. Its form also resembles that of Lagrangians commonly used in biophysics~\cite{Bialek2012} to optimize MI subject to constraints, such as those in rate-distortion theory~\cite{Cover1991} and the information bottleneck method~\cite{tishby2000information}.  

MI $I(X;Z)$, in Eq.~(\ref{eqm1}), is defined as
\begin{equation}
    I(X;Z) := \int dx dz P_{X,Z}(x,z) \log_2 \left[ \frac{P_{X,Z}(x,z)}{P_X(x)P_Z(z)}
    \right],
    \label{eqm2}
\end{equation}

\noindent where $P_{X,Z}(x,z)$ is the joint distribution of the input and output variables, and $P_X(x) = \int dz P_{X,Z}(x,z)$ and $P_Z(z) = \int dx P_{X,Z}(x,z)$ are the corresponding marginal distributions. The lowercase letters  $x$ and $z$ are used to represent the input and output states, respectively, in Euclidean space. MI is measured in \textit{bits} due to base 2 in the logarithm. The 2-WD $W(X,Z)$, a geometric measure of dissimilarity between the input distribution $P_X(x)$ and the output distribution $P_Z(z)$, is defined as
\begin{eqnarray}
W(X,Z) &:=& \sqrt{ \min_{\pi(x,z) \in \bm{\Pi}(P_X, P_Z)} \int dx dz \left(x-z\right)^2 \pi (x,z)},
\label{eqm3}
\end{eqnarray}

\noindent where $\bm{\Pi}(P_X, P_Z)$ denotes the set of all joint distributions $\pi(x,z)$ such that the marginals satisfy: $P_X(x)=\int dz \pi(x,z)$ and $P_Z(z)=\int dx \pi(x,z)$ with $\pi(x,z) \ge 0$. $\pi(x,z)$ represents a transport plan. Here, $\left(x-z\right)^2$ quantifies the transport cost between two points, and its expected value under the distribution $\pi(x,z)$ represents the total cost of the transport plan. The 2-WD, thus, compares the marginal distributions through the minimum transport cost required to transform one distribution into the other. It is a non-negative quantity that yields zero when $P_X$ and $P_Z$ are equal, and satisfies the axioms of a metric~\cite{Villani2008}. It also provides differential geometric and thermodynamic perspectives on the dynamics of the Fokker-Planck equation~\cite{Villani2008,jordan1998variational,ito2024geometric,oikawa2025experimentally}, and these features make it well-suited to the dynamics of the chemical Langevin equation. Because W(X,Z) has the units of $X$ or $Z$, the variables must be expressed on comparable scales before computing. A detailed interpretation of $W(X,Z)=1$ with appropriate units and biological implications is discussed in the \textit{Supplementary Sec.~S4}.

Note that Kullback-Leibler divergence (KLD) is also widely used to compare distributions. However, to quantify distributional mismatch, we use the 2-WD because it measures geometric displacement of probability mass and provides a symmetric measure of shifts, broadening, compression, or redistribution of probability. In contrast, KLD measures relative density mismatch and is asymmetric in nature. The use of the 2-WD is further supported by observations from Wasserstein GANs \cite{Arjovsky2017}. Gaussian limiting cases for the marginal distributions comparing KLD and 2-WD are discussed in the \textit{Supplementary Sec.~S5}.

\begin{figure*}[!t]
\centering
\includegraphics[width=0.95\columnwidth,angle=0]{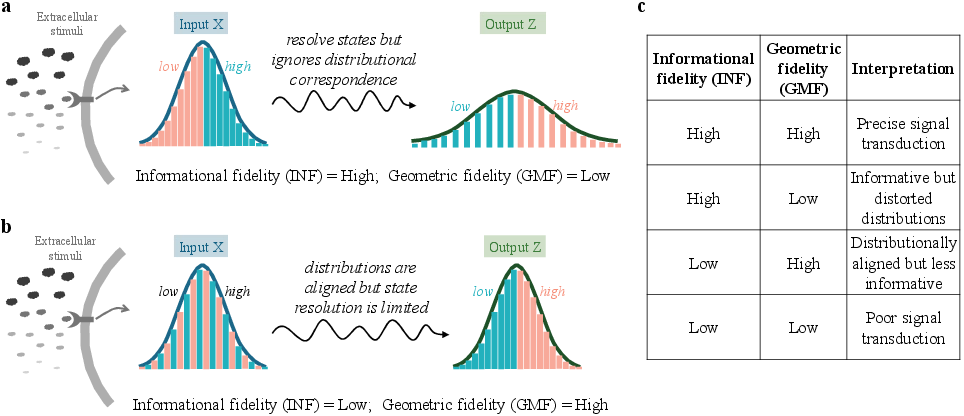}
\caption{\textbf{Schematic illustrations of informational and geometric fidelity.}
\textbf{a} Informational fidelity (INF) characterizes state-wise mapping between input and output distributions. This mapping resolves states accurately but sacrifices distributional correspondence.
\textbf{b} Geometric fidelity (GMF) accounts for shape, scale (variability) and mean-wise mapping of input distribution to output. While this geometric fidelity cannot measure the precise discrimination of input states in the output, it does measure their overall distributional correspondence.
\textbf{c} The two fidelities, together, can characterize different signaling behaviors. The table summarizes the various combinations of these two fidelities and their interpretation in signaling.
The colored bars in panels \textbf{a} and \textbf{b} denote the representative states associated with input and output.}
\label{f1a}
\end{figure*}

\noindent \textbf{Significance of dual-fidelity.} High INF corresponds to accurate state discrimination, where the output reliably distinguishes different input states (e.g., \textit{low} vs. \textit{high}) (Fig.~\ref{f1a}a), but does not quantify whether the output preserves the distributional structure of the input. In contrast, high GMF corresponds to preservation of input-distribution features in the output, including mean, variability, and overall shape, but does not by itself ensure precise input-state discrimination  (Fig.~\ref{f1a}b). Using an illustrative example, we show in \textit{Supplementary Sec.~S6} that two outputs can carry similar INF about the same unimodal input, while GMF clearly distinguishes whether the output remains unimodal or splits into a bimodal distribution.

Since the two fidelities capture distinct aspects in signal transduction, we propose the dual-fidelity framework as a means to characterize the distinct strategies that biochemical systems can exhibit for signaling. To interpret this interplay, we summarize how different combinations of the two fidelities correspond to distinct signaling behaviors (see Fig.~\ref{f1a}c). When both INF and GMF are high, signaling is precise, combining state discrimination with distributional correspondence. High INF but low GMF give informative signaling, where input states are distinguishable but the output distribution is distorted. High GMF but low INF gives geometric signaling, where distributional correspondence is maintained, but state discrimination is weak. When both are low, signaling is poor. However, all four combinations can, in principle, arise depending on the specific biological system and signaling context. This assessment indicates that neither fidelity alone provides a complete description of signaling; together, they can capture the complementary dimensions necessary to elucidate the efficacy of signal transmission. To illustrate this framework, we next analyze canonical gene regulatory motifs.

\begin{figure*}[!t]
\centering
\includegraphics[width=0.9\columnwidth,angle=0]{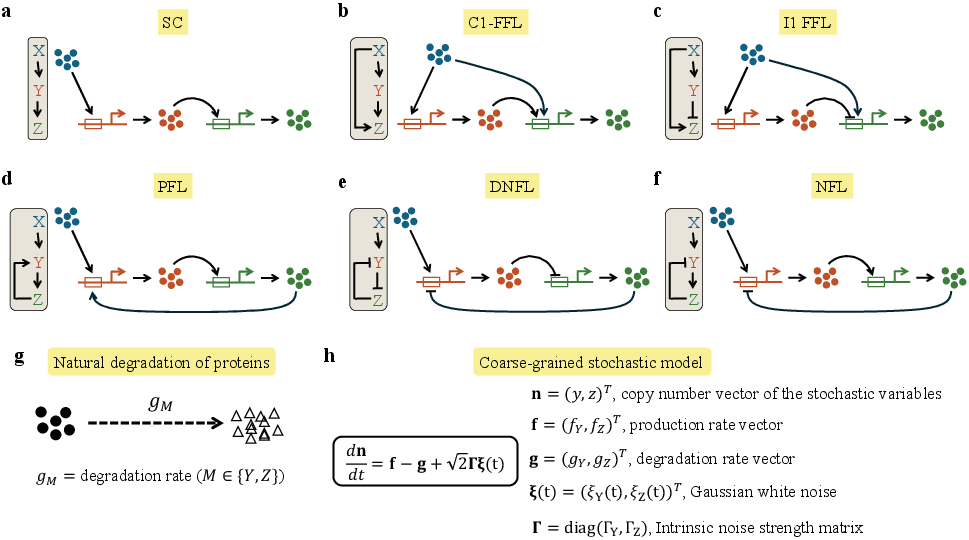}
\caption{\textbf{Schematics of gene regulatory motifs and governing dynamical equation}.
\textbf{a} Simple cascade: the input $X$ activates $Y$, which subsequently activates $Z$. \textbf{b} Coherent type-1 feed-forward loop: $X$ activates both $Y$ and $Z$, and $Y$ activates $Z$. \textbf{c} Incoherent type-1 feed-forward loop: $X$ activates both $Y$ and $Z$, but $Y$ represses $Z$. \textbf{d} Positive feedback loop: $X$ activates $Y$, which again activates $Z$, and $Z$ activates $Y$ forming the feedback. \textbf{e} Double negative feedback loop: $X$ activates $Y$, which represses $Z$, and $Z$ represses $Y$. \textbf{f} Negative feedback loop: $X$ activates $Y$, which again activates $Z$, but $Z$ represses $Y$.
\textbf{g} Natural degradation with rate $g_M$ ($M \in \{Y, Z \}$) is illustrated.
\textbf{h} The stochastic model for the gene regulatory motifs is shown. Here, $\bm{\xi}(t)$ denotes the Gaussian white noise satisfying $\overline{\xi_M(t)}=0$ and $\overline{\xi_M(t)\xi_{M'}(t')}=\delta_{M{M'}}\delta(t-t')$ with $M \in \{ Y,Z \}$ and $M' \in \{ Y,Z \}$. The overbar $\overline{\cdots}$ represents the ensemble average.}
\label{f2}
\end{figure*}

\noindent \textbf{Gene regulation and fidelity optimization.} We apply the dual-fidelity framework to six canonical regulatory motifs -- simple cascade (SC), coherent type-1 feed-forward loop (C1-FFL), incoherent type-1 feed-forward loop (I1-FFL), positive feedback loop (PFL), double negative feedback loop (DNFL), and negative feedback loop (NFL) (Figs.~\ref{f2}a–f). In all cases, proteins degrade at rates $\bm{g}$ (Fig.~\ref{f2}g) and are produced according to regulatory functions $\bm{f}$ (Fig.~\ref{f2}h).

Stochastic dynamics of the motifs are modeled using coupled differential equations based on the Langevin formalism (see Fig.~\ref{f2}h and \hyperref[sec4]{Materials and methods}). Solving these equations using linear noise approximation (LNA)~\cite{Kampen2007, Gardiner2009} yields the following analytical forms for the intensity of the output noise $\eta_Z^2$ and the input–output covariance $\zeta_{XZ}$ in the steady-state,  
\begin{eqnarray}
    \eta_Z^2 &=& \frac{1}{\mu_Z}+\Phi_1+\Phi_2 \eta_X^2,
    \label{eqm3a}
    \\
    \zeta_{XZ} &=& \Psi \eta_X^2,
    \label{eqm3b}
\end{eqnarray}

\noindent where $\eta_X^2$ is the intensity of the input noise. The coefficients $\Phi_1$, $\Phi_2$, and $\Psi$ depend on system parameters and regulatory sensitivities $f^\prime_{ML} = (\partial f_M / \partial n_L)$ in the steady-state, where $f_M$ (an element of $\bm{f}$) is the production rate in the system $M \in \{Y,Z\}$, and $n_L \in \{x,y,z\}$ is a molecule count (copy number) of the system $L \in \{X,Y,Z\}$. Here $f^\prime_{ML}$ denotes how strongly a gene encoding $M$ is regulated by $L$. The motif-specific forms of $f_M$ and $f^\prime_{ML}$ are provided in the \textit{Supplementary Table~S1}.

\begin{figure*}[!t]
\centering
\includegraphics[width=0.9\columnwidth,angle=0]{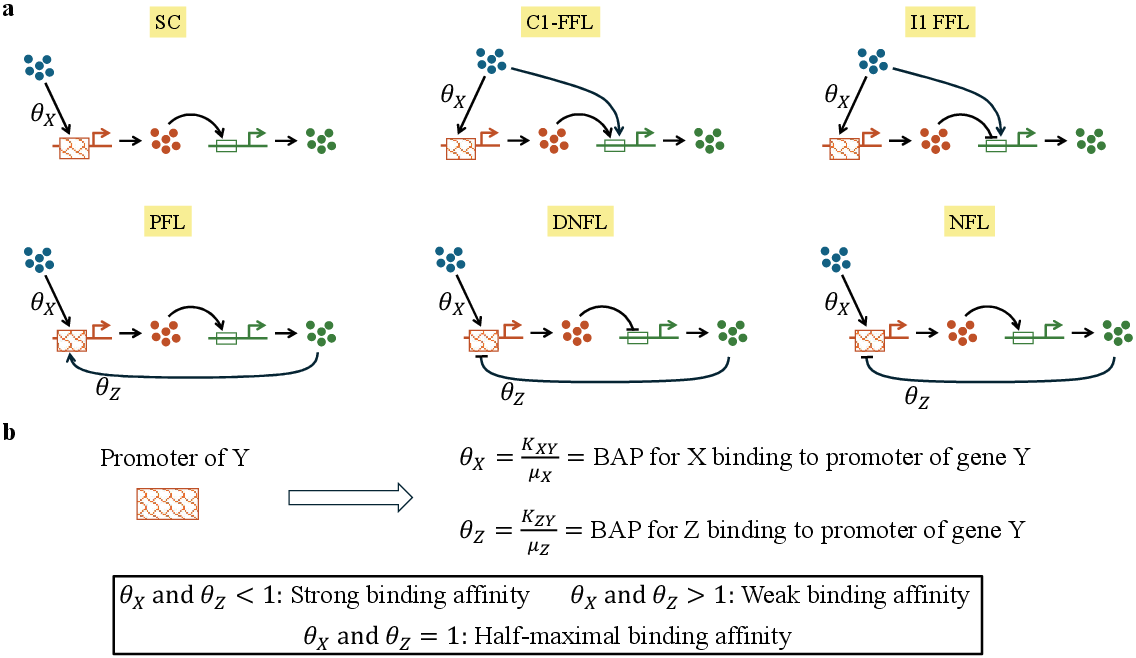}
\caption{\textbf{Binding affinity parameters $\theta_X$ and $\theta_Z$.}  
\textbf{a} Schematic illustration of $\theta_X$ and $\theta_Z$ across different network motifs.  
\textbf{b} Definition of the binding affinity parameters. $\theta_X$ and $\theta_Z$ represent the effective binding affinities of $X$ and $Z$, respectively, to the promoter of gene $Y$.}
\label{f3}
\end{figure*}

Among the regulatory sensitivities, we highlight $f^\prime_{YX}$ and $f^\prime_{YZ}$, which describe how $X$ and $Z$ regulate the promoter of gene $Y$. We define the corresponding binding affinity parameters (BAPs) as $\theta_X=K_{XY}/\mu_X$ and $\theta_Z=K_{ZY}/\mu_Z$ (Fig.~\ref{f3}a), where $\mu_X$ and $\mu_Z$ denote the steady-state mean copy number of $X$ and $Z$, respectively. The parameter $\theta_Z$ appears only in feedback motifs, where $Z$ regulates $Y$. Values $\theta_X<1$ ($\theta_Z<1$), $\theta_X>1$ ($\theta_Z>1$), and $\theta_X=1$ ($\theta_Z=1$) represent strong, weak, and half-maximal effective binding, respectively (Fig.~\ref{f3}b). See \textit{Supplementary Sec.~S2} for detailed discussions.

To facilitate analytical tractability, we adopt Gaussian channel approximations for both MI and the 2-WD (see Eqs.~(\ref{eqm6})~and~(\ref{eqm8})), allowing INF and GMF to be expressed in terms of the motif noise characteristics. We fix all other biochemical parameters and vary the BAPs together with $\lambda$. For $\theta' \in \{\theta_X,\theta_Z \}$, we consider three regimes: $\theta'<1$, $\theta'=1$, and $\theta'>1$, represented by the values $\theta'=0.5$, $\theta'=1$, and $\theta'=2$, respectively. All nine pairwise combinations of $(\theta_X,\theta_Z)$ are analyzed for the PFL, DNFL, and NFL motifs, while for SC, C1-FFL, and I1-FFL, only $\theta_X$ is varied. The parameter $\lambda$ can be sampled from a discrete set. With this convention, Eq.~(\ref{eqm1}) takes the parameterized form $\mathcal{L}(\eta_X^2;\theta_X,\theta_Z,\lambda)$ when the set of parameters $(\eta_X^2, \theta_X,\theta_Z,\lambda)$ is fixed. For varying combinations of $(\theta_X,\theta_Z)$ and different values of $\lambda$, we optimize the Lagrangian $\mathcal{L}$ in Eq.~(\ref{eqm1}) with respect to $\eta_X^2$ within the Gaussian framework, which yields the optimal input noise $\eta_X^{2*}$, via
\begin{eqnarray}
    \eta_X^{2*} &=& \arg \max_{\eta_X^2} \mathcal{L}(\eta_X^2;\theta_X,\theta_Z,\lambda),
    \label{eqm4}
\end{eqnarray}

\noindent The corresponding INF and GMF are then evaluated using Eqs.~(\ref{eqm6})~and~(\ref{eqm8}) at this optimized operating point. We note that Eq.~(\ref{eqm4}) applies to both the set of motifs $\{\text{SC, C1-FFL, I1-FFL}\}$ and $\{\text{PFL, DNFL, NFL}\}$, where $\theta_Z$ dependence is absent for the former set (see Fig.~\ref{f3}a).

\begin{figure*}[!t]
\centering
\includegraphics[width=0.9\columnwidth,angle=0]{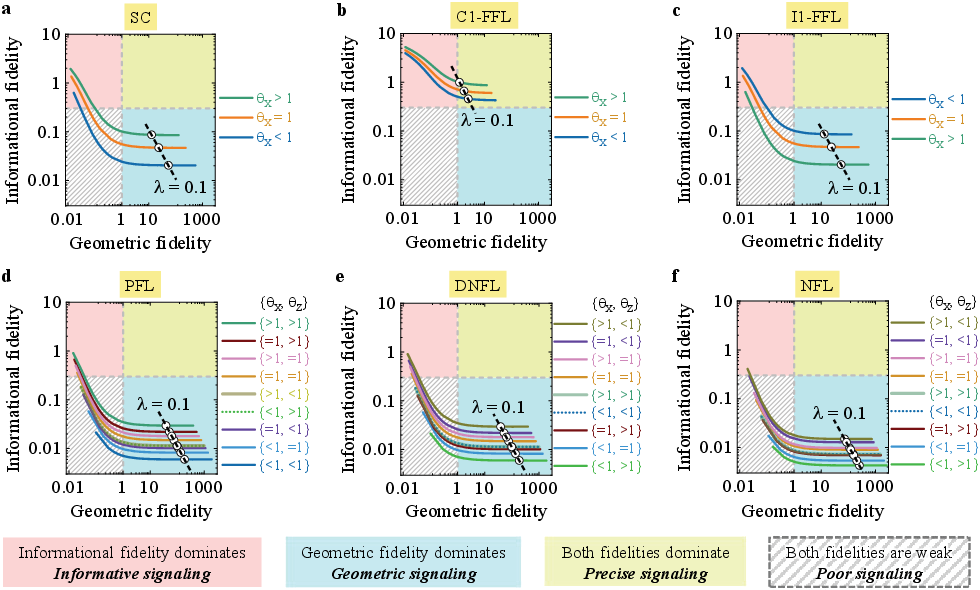}
\caption{\textbf{Interplay between informational and geometric fidelities across network motifs.}  
The relationships between the two fidelities obtained from the dual-fidelity optimization framework for six canonical motifs:  
\textbf{a} Simple cascade (SC),  
\textbf{b} Coherent type-1 feed-forward loop (C1-FFL),  
\textbf{c} Incoherent type-1 feed-forward loop (I1-FFL),  
\textbf{d} Positive feedback loop (PFL),  
\textbf{e} Double negative feedback loop (DNFL), and  
\textbf{f} Negative feedback loop (NFL).  
Each curve represents how the operating regime shifts as BAP and $\lambda$ are varied. 
BAPs are fixed, while $\lambda$ varies along each curve. Open circles mark $\lambda=0.1$, and black dashed lines connect these points to represent the variations in INF and GMF for a fixed $\lambda$ and varying BAP. Pink, cyan, pale yellow-green, and grey-hatched domains in each panel denote informative, geometric, precise, and poor signaling regimes, respectively (also see Fig.~\ref{f1a}c). These color-coded regimes are defined at the bottom of the figure. Grey dashed horizontal and vertical lines mark the illustrative thresholds $I(X;Z)=0.3$ bits and $W(X,Z)^{-1}=1\,(\mathrm{copy\ number})^{-1}$, respectively, separating these signaling regimes.
The model parameters used to generate these plots are mentioned in Supplementary Information Sec.~S3.
}
\label{f4}
\end{figure*}

\noindent \textbf{Dual-fidelity behavior due to differential binding affinities.} Using Eq.~(\ref{eqm4}), we compute the optimal input noise $\eta_X^{2*}$ for each motif in Figs.~\ref{f2}a-f across different combinations of BAPs and different values of $\lambda$. The corresponding optimal output noise $\eta_Z^{2*}$ and input-output covariance $\zeta_{XZ}^*$ follow from Eqs.~(\ref{eqm3a})–(\ref{eqm3b}), which determine MI and the 2-WD via Eqs.~(\ref{eqm6})~and~(\ref{eqm8}). In these calculations, we fix the mean copy numbers of $X$, $Y$, and $Z$ to be equal (see \textit{Supplementary Sec.~S3}) to isolate the effect of copy-number variability. The resulting interplay between INF and GMF is shown in Fig.~\ref{f4}, with each curve representing a specific binding affinity profile over varying $\lambda$.

We now analyze the effect of binding affinity at $\lambda = 0.1$ across all motifs. In SC, weak binding of $X$ ($\theta_X > 1$) increases INF but decreases GMF, while strong binding ($\theta_X < 1$) produces the opposite trend (Fig.~\ref{f4}a). Weak binding increases sensitivity to fluctuations, raising input-output covariance ($|\zeta_{XZ}|$) and INF, but also increases output noise and lowers GMF (as is directly evident from Eqs.~(\ref{eqm6})~and~(\ref{eqm8})). Strong binding saturates the promoter, damping output variability and improving GMF at the cost of INF. C1-FFL shows a similar trend, although with different magnitudes due to its coherent two-arm activation of $Z$ \cite{Mangan2003, Alon2006} (Fig.~\ref{f4}b).

I1-FFL shows an inverted pattern (Fig.~\ref{f4}c) because its activating and repressing arms oppose each other \cite{Mangan2006}, and strong binding in one arm can offset fluctuations in the other, reducing variability without saturation. PFL follows SC/C1-FFL: weak binding of both $X$ and $Z$ ($\theta_X > 1$, $\theta_Z > 1$) yields high INF but low GMF, while strong binding of both does the reverse (Fig.~\ref{f4}d). DNFL and NFL behave similarly to each other but differ from the other motifs. In both, weak $X$ binding combined with strong $Z$ binding favors INF, whereas strong $X$ binding combined with weak $Z$ binding favors GMF (Fig.~\ref{f4}e,f).

Overall, these results highlight a trade-off between the two fidelities that is shaped by how binding affinities of $Y$'s promoter are modulated across different network topologies. Coherent architecture tends to align in its response to affinity changes, whereas incoherent and feedback-containing motifs can invert or alter this trend due to opposing regulatory effects. Such patterns suggest that binding affinity can act as a biochemical lever for tuning the balance between INF and GMF. Next, we examine the relative motif-specific preferences for the two fidelities.

\noindent \textbf{Motif-specific patterns in dual-fidelity space.} To compare motif performances, we map the dual-fidelity space into four qualitative regimes: informative, geometric, precise, and poor signaling. The regime boundaries are chosen for interpretability and are not strict cutoffs. We choose $I(X;Z) \ge 0.3$ bits as the minimum for biologically meaningful state discrimination \cite{Cheong2011}, and $W(X,Z)^{-1} \ge 1$ in (copy numbers)$^{-1}$ unit as the minimum for reliable distributional correspondence (explicitly discussed in \textit{Supplementary Sec.~S4}). 

A motif-specific trade-off between the two fidelities is observed as a function of $\lambda$ (see Fig.~\ref{f4}). In the dual-fidelity space, SC and I1-FFL span a broad range, accessing informative, geometric, and poor signaling regimes (Figs.~\ref{f4}a,c). C1-FFL is largely confined to informative and precise regimes (Fig.~\ref{f4}b), consistent with its role in sustaining coordinated responses for robust signal transmission \cite{Mangan2003, Alon2006}. Feedback motifs (PFL, DNFL, NFL) primarily explore geometric and poor signaling regimes, with limited access to informative signaling. Among them, NFL shows the strongest bias toward GMF (Figs.~\ref{f4}d–f). These trends highlight how network structure biases the trade-offs observed between state resolution and distributional correspondence. We note that the realization of these trade-offs depends on how the underlying biophysical properties of each motif, including the BAPs ($\theta$) and network topology, constrain the joint variation of INF and GMF. We emphasize that the observed trade-off is not a trivial consequence of the Lagrangian formulation in Eq.~\ref{eqm1}, as demonstrated through a limiting case in \hyperref[sec4]{Materials and methods}.

These motif-specific preferences are broadly consistent with known motif abundance in transcription networks. C1-FFL and I1-FFL are frequent in bacterial and yeast networks \cite{Shen-Orr2002, Mangan2006, Alon2006, Alon2007}, and our results suggest two possible functional interpretations: C1-FFL can access precise signaling regimes, whereas I1-FFL provides flexible tuning across fidelity modes. Feedback motifs, especially NFL, are biased toward GMF, consistent with a possible role in stabilizing responses rather than enhancing state discrimination. These observations suggest that motif topology and binding affinity tune where a network operates in the dual-fidelity landscape.

\begin{figure*}[!t]
\centering
\includegraphics[width=0.8\columnwidth,angle=0]{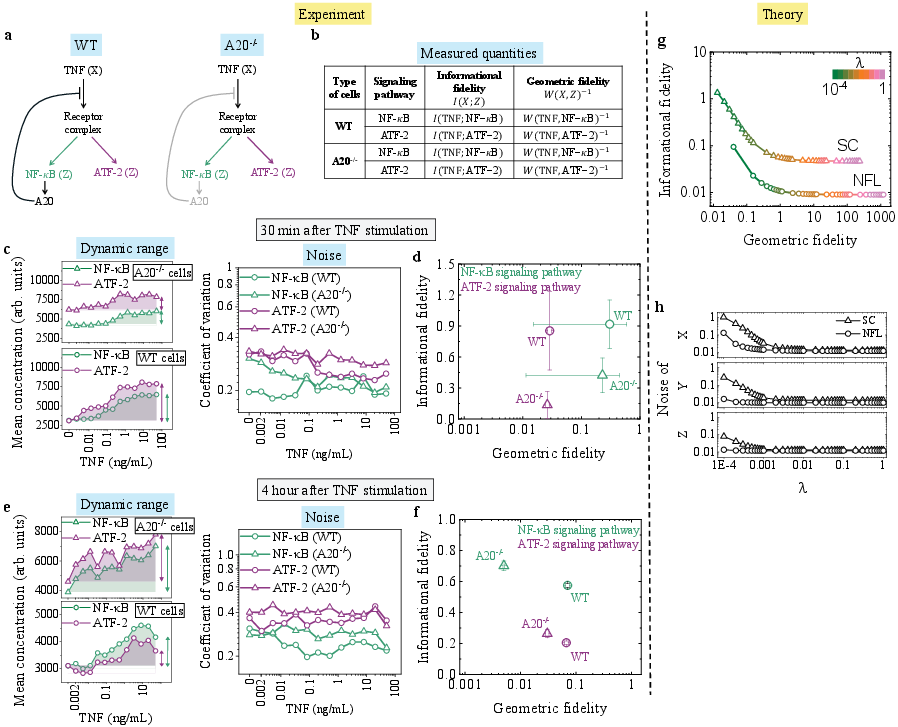}
\caption{\textbf{Experimental and theoretical signatures of dual-fidelity in TNF signaling}  
\textbf{a} Schematic of TNF signaling in WT and A20$^{-/-}$ cells. The grey arrow in A20$^{-/-}$ cells means that the negative feedback is absent. Here, green and purple labels are used to distinguish between NF-$\kappa$B and ATF-2 pathways.
\textbf{b} Summary of measured quantities: INF $I(X;Z)$ and geometric fidelity $W(X,Z)^{-1}$ for NF-$\kappa$B and ATF-2 pathways in both cell types.
\textbf{c} Mean concentration and CV as functions of TNF concentration at 30 minutes after stimulation. These profiles characterize the dynamic range and noise (cell-to-cell variability) of each output, and are generated from the reported dose-response statistics \cite{Cheong2011}. Open circles denote WT cells, whereas open triangles represent A20$^{-/-}$ cells. See Supplementary Information Sec.~S7 for details.
\textbf{d} Informational fidelity plotted against geometric fidelity at 30 minutes after TNF stimulation. Informational fidelity is expressed in bits, and geometric fidelity is expressed in the unit reciprocal to TNF concentration.
\textbf{e} Mean concentration and CV as functions of TNF concentration at 4 hours after stimulation.
\textbf{f} Informational fidelity plotted against geometric fidelity at 4 hours after TNF stimulation.
\textbf{g} Dual-fidelity curves for simple cascade (SC) with $\theta_X=1$ and negative feedback loop with $\theta_X=1$ and $\theta_Z=1$, across different values of $\lambda$. These plots are reproduced from Figs.~\ref{f4}a~and~\ref{f4}f. Here, the informational fidelity is also expressed in bits, but the geometric fidelity has the unit reciprocal to the copy numbers (units of $X$ and $Z$). Here, the solid lines are drawn by connecting the symbols (circle/triangle) to show the variations clearly. Each curve contains a color gradient defined by the change in $\lambda$ values.
\textbf{h} Optimal noise ($\eta_L^{2*}$) of $X$, $Y$, and $Z$ for SC and negative feedback loop (NFL) motifs with similar BAPs. $\eta_X^{2*}$ is obtained from Eq.~(\ref{eq4}), while the other two are computed from their expressions in terms of the noise of $X$, as provided in the Supplementary Information Sec.~S1. Here, the solid lines are drawn by connecting the symbols (circle/triangle) to show the variation clearly.
In panels \textbf{c} and \textbf{e}, green and purple shaded areas indicate the dynamic ranges of NF-$\kappa$B and ATF-2, respectively, for each cell type.
In panels \textbf{d} and \textbf{f}, symbols show INF and GMF values estimated considering all 13 TNF dose values for each pathway, cell type, and time point. Reported cell counts and experimental replicate identifiers are not used for this reanalysis. The vertical error bars show one bootstrap standard deviation of INF. Horizontal error bars show GMF uncertainty obtained by first-order propagation, $s_{\mathrm{GMF}}=s_W/W^2$, where $s_W$ is the bootstrap standard deviation of the 2-WD. Bootstrap standard deviations are calculated from valid estimates obtained in 100 resampling attempts per dataset.}
\label{f5}
\end{figure*}

\noindent \textbf{Dual-fidelity regulation in TNF signaling.} We next examine whether signaling systems exhibit experimentally detectable signatures of dual-fidelity regulation. To address this, we analyze published single-cell data of nuclear factor-$\kappa$B (NF-$\kappa$B) and activating transcription factor-2 (ATF-2) responses to TNF stimulation in wild-type (WT) and A20-deficient mutant (A20$^{-/-}$) mouse fibroblast cells (Fig.~\ref{f5}a), as reported by Cheong \textit{et al.} \cite{Cheong2011}. In WT cells, A20 mediates a negative feedback to the TNF pathway, whereas this regulation is absent in the A20$^{-/-}$ mutant. We assign TNF as the input $X$ and either NF-$\kappa$B or ATF-2 as the output $Z$, depending on the pathway analyzed. We quantify the mean nuclear concentration of NF-$\kappa$B and ATF-2, and the corresponding coefficient of variations (CVs) to characterize the dynamic range and noise properties (cell-to-cell variability), respectively. We then compute the INF and GMF, as summarized in Fig.~\ref{f5}b. See \hyperref[sec4]{Materials and methods} for further details.

To examine how A20-mediated regulation affects signaling over time, we analyze the dynamic range, noise, and dual-fidelities at 30 minutes and 4 hours after TNF stimulation. The dynamic range is measured as the difference in mean concentrations of NF-$\kappa$B (ATF-2) between the highest and lowest TNF concentrations. This quantity captures the effective response span of each biochemical node over the measured TNF range. A reduced dynamic range or a plateauing dose-response profile indicates response compression or saturation-like behavior, where further increases in TNF produce only limited changes in the output. Such node-level biochemical properties directly influence the fidelity measures: a broader dynamic range can improve the distinguishability of TNF levels and thereby increase INF. In contrast, lower cell-to-cell variability and a more constrained response distribution can reduce input-output distributional mismatch and thereby increase GMF. Thus, the fidelity patterns are linked to the biochemical properties of each output node as well as to the feedback topology of the network. 

At 30 minutes, NF-$\kappa$B (ATF-2) in WT cells retains a larger dynamic range than in A20$^{-/-}$ cells, whereas cell-to-cell variability is not strongly separated (Fig.~\ref{f5}c). This gives higher INF for both NF-$\kappa$B  and ATF-2 in WT cells, while the GMF remains comparable for both cell types (Fig.~\ref{f5}d). At 4 hours, this behavior changes, as A20$^{-/-}$ spans a broader output range and shows larger variability, which increases INF but reduces GMF. In contrast, the WT cells become more constrained and less variable, lowering INF but increasing GMF (Fig.~\ref{f5}e,f). These behaviors are consistent with the time-dependent engagement of A20-mediated negative feedback. In the early phase, the feedback is not sufficiently effective to generate a strong difference in cell-to-cell variability, and the two cell types mainly differ in dynamic range. In the late phase, active A20-mediated feedback suppresses the WT response range and constrains response variability relative to A20$^{-/-}$ mutants. Although ATF-2 is not directly regulated by A20, it shows a similar but weaker trend, because both NF-$\kappa$B and ATF-2 share the upstream TNF-receptor complex.

Importantly, these results highlight the limitations of interpreting signal transmission solely through INF. An INF-only view would suggest that WT cells perform better in the early phase, whereas A20$^{-/-}$ cells perform better in the late phase. This temporal inversion shows that INF captures time-dependent state discrimination, but does not by itself explain the regulatory role of A20-mediated feedback. This issue becomes most relevant in the late phase, when the higher INF of A20$^{-/-}$ cells may appear to suggest superior signaling. GMF, here, offers a complementary view, showing that WT cells exhibit closer input-output distributional correspondence when feedback is engaged, despite lower INF. This behavior highlights the role of A20-mediated feedback in differentially modulating state discrimination and distributional correspondence in the present TNF-signaling setup.

The separation of dual fidelities between WT and A20$^{-/-}$ cells in the late phase (Fig.~\ref{f5}f) qualitatively follows the trends predicted from the theoretical analysis. To illustrate this correspondence, we reproduced the theoretical curves in Fig.~\ref{f5}g from Figs.~\ref{f4}a~and~\ref{f4}f for SC (with $\theta_X = 1$), which lacks feedback, and NFL (with $\theta_X = 1$ and $\theta_Z = 1$), which contains negative feedback. We note that in the experiment, GMF is expressed in the unit reciprocal to TNF concentration, while in the theory, it is expressed in the unit reciprocal to copy number. Hence, absolute magnitudes are not directly comparable; rather, we compare the trends and separations. Although the model topologies differ from the biochemical details of the TNF--NF-$\kappa$B/ATF-2 network, both the experimental and theoretical results display a consistent qualitative picture -- negative feedback shifts signaling toward higher GMF and lower INF, whereas the absence of feedback produces the opposite trend. In the theoretical analysis, this fidelity shift arises because the NFL suppresses noise in all gene products ($X$, $Y$, and $Z$) relative to SC (Fig.~\ref{f5}h).

\begin{figure*}[!t]
\centering
\includegraphics[width=0.8\columnwidth,angle=0]{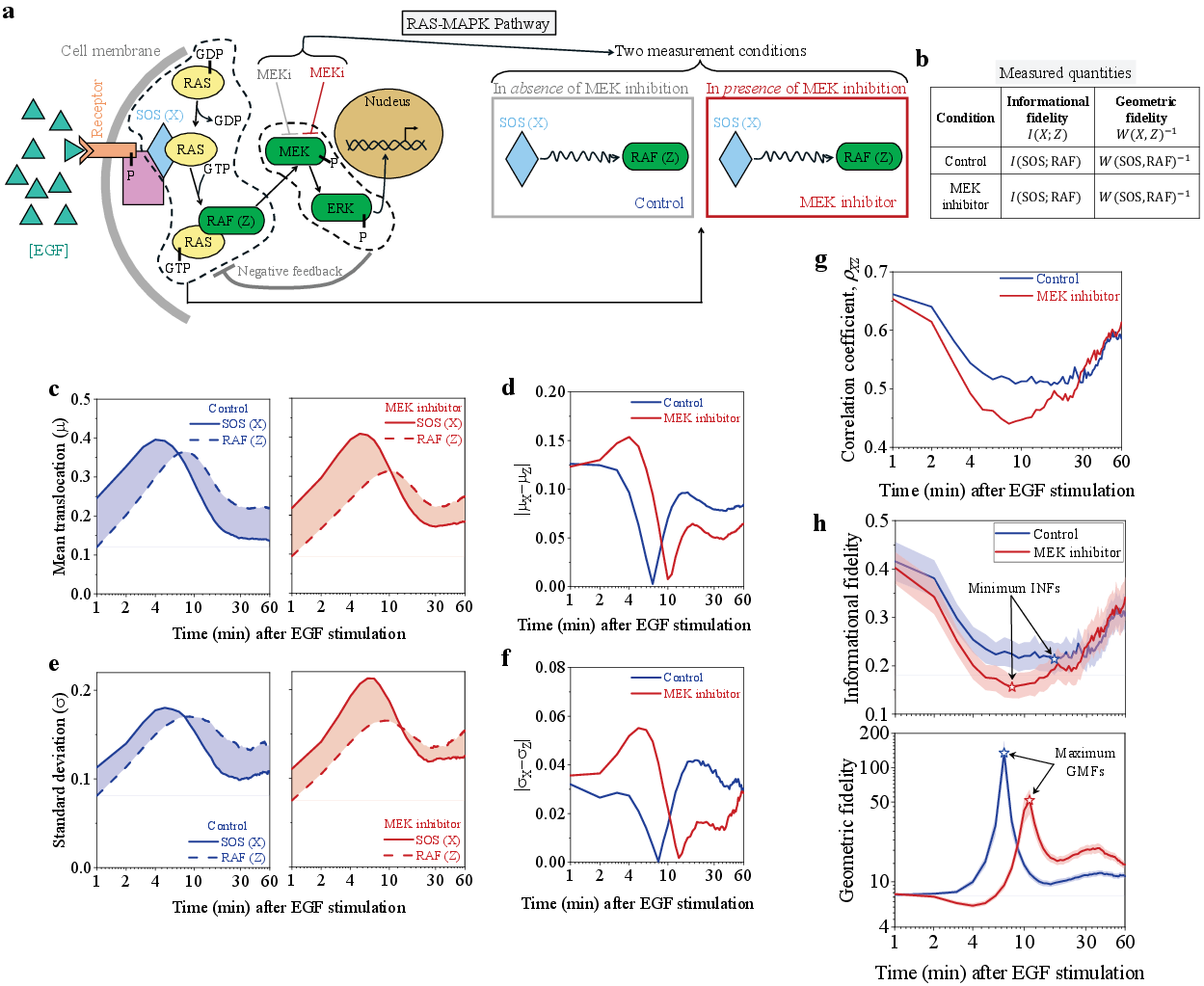}
\caption{
\textbf{Dual-fidelity signature in RAS-MAPK signaling}  
\textbf{a} Schematic of the RAS-MAPK signaling pathway. The analysis compares two experimentally probed conditions: control and MEK-inhibited conditions. We use blue and red labels to denote the control and MEK-inhibited conditions in the following panels.
\textbf{b} Summary of measured quantities: informational fidelity $I(X;Z)$ and geometric fidelity $W(X,Z)^{-1}$ for both conditions.
\textbf{c} Temporal variations in mean translocation levels for SOS (solid) and RAF (dashed) under control and MEK-inhibited conditions.
\textbf{d} Absolute difference between the mean translocation levels of SOS and RAF for both conditions.
\textbf{e} Temporal variation in the corresponding standard deviations of SOS (solid) and RAF (dashed) translocation.
\textbf{f} Absolute difference between the standard deviations of SOS and RAF for both conditions. The mean and standard deviation profiles were computed from the reported single-cell time-series data \cite{Umeki2025}.
\textbf{g} Temporal variation in the correlation coefficient between SOS and RAF under both conditions.
\textbf{h} Temporal profiles of informational fidelity and geometric fidelity between SOS and RAF under control and MEKi conditions. The marked points indicate the minimum of the INF trajectory and the maximum of the GMF trajectory for both conditions, which summarize the strongest transient loss of statistical coupling and the strongest transient distributional correspondence during each signaling trajectory. Informational fidelity is expressed in bits, whereas geometric fidelity is dimensionless because the reported fluorescence intensities are normalized values \cite{Umeki2025}. 
In panels \textbf{c} and \textbf{e}, curves show arithmetic means and sample standard deviations
across cells (cell-to-cell variability), respectively. Shaded bands indicate the region between the SOS and RAF curves. In panel \textbf{h}, the central solid lines represent the estimated INF and GMF from the data. Successive time points represent repeated measurements of the same cells. However, the shaded bands show the point estimate plus or minus a symmetric bootstrap error obtained from 500 paired resamples per condition and time point. The error magnitude is the smaller non-negative distance from the point estimate to the 16th and 84th bootstrap percentiles, capped to prevent a negative lower bound. All statistics in panels \textbf{c}-\textbf{h} use 744 cells under control conditions and 706 cells under MEK inhibition at each time point. 
}
\label{f6}
\end{figure*}

\noindent \textbf{Dual-fidelity regulation in RAS-MAPK signaling.} To examine the experimental signature of the dual-fidelity in an intracellular setting, where both the input and output are dynamically varying cellular components, we study the RAS--mitogen-activated protein kinase (RAS-MAPK) signaling pathway (Fig.~\ref{f6}a). We analyze published single-cell data of epidermal growth factor (EGF)-stimulated SOS and RAF membrane translocation, under control and MEK-inhibited conditions, as reported by Umeki \textit{et al.} \cite{Umeki2025}. MEK inhibitor (MEKi) treatment suppresses MEK activity and thereby reduces downstream ERK activation. We assign SOS as the input $X$ and RAF as the output $Z$, and quantify the biochemical properties, such as mean translocation, standard deviation characterizing cell-to-cell variability, and SOS-RAF correlation coefficient from the reported time-series data. We also quantify the INF and GMF for both conditions, as summarized in Fig.~\ref{f6}b, using the Gaussian-approximated expressions in Eqs.~(\ref{eqm6})~and~(\ref{eqm7}). See \hyperref[sec4]{Materials and methods} for further details.

The temporal profiles of SOS and RAF mean translocation ($\mu_X$ and $\mu_Z$), cell-to-cell variability ($\sigma_X$ and $\sigma_Z$), and correlation ($\rho_{XZ}$) show non-monotonic behavior under both control and MEKi conditions (Fig.~\ref{f6}c,e,g). The absolute difference in mean ($\lvert\mu_X-\mu_Z\rvert$) and cell-to-cell variability ($\lvert\sigma_X-\sigma_Z\rvert$) also vary non-monotonically, with noticeable inversions and temporal lags between the two conditions (Fig.~\ref{f6}d,f). These behaviors are consistent with the temporal organization of RAS-MAPK signaling. Specifically, MEKi can perturb several negative feedback interactions from the MEK-ERK module to the SOS-RAF module and its upstream, as reported in previous studies \cite{Corbalan-Garcia1996, Saha2012, Umeki2025}. In the control condition, negative feedback stabilizes SOS and RAF dynamics, thereby supporting coordinated signal transmission. However, MEKi perturbs these feedback interactions owing to time-dependent accumulation of MEK/ERK. Moreover, the activity of the EGF receptor also changes over time after EGF stimulation, which further affects the SOS and RAF activities. Together, the time-dependent EGF response and MEKi-induced feedback perturbation provide an explanation for the observed temporal variations in the biochemical properties.

The behavior of these biochemical properties directly explains the temporal profiles of INF and GMF for the two conditions (Fig.~\ref{f6}h). Specifically, the correlation coefficient, $\rho_{XZ}$, controls the nature of INF, whereas $\lvert\mu_X-\mu_Z\rvert$ and $\lvert\sigma_X-\sigma_Z\rvert$ together determine the nature of GMF. Both fidelities also exhibit the temporal inversions and lags between the two conditions due to the interplay of time-dependent EGF response and MEKi-induced feedback perturbation. For both conditions, INFs pass through a minimum and the GMFs pass through a maximum (see Fig.~\ref{f6}h). Although these extreme points appear at different times along the trajectories, they can provide a qualitative signature of broader biophysical principles governing the RAS-MAPK signaling.

To be specific, an analysis based solely on the minimum INF would identify the strongest transient loss of SOS-RAF statistical coupling under each condition, as a result of which the signaling seems imprecise at the operating time points. However, GMF provides a complementary interpretation by identifying the strongest transient distributional correspondence between SOS and RAF, leading to geometrically precise signaling. Moreover, under MEKi, both the minimum INFs and maximum GMFs are reduced compared to the control condition, indicating poor signal transmission at the corresponding operating time points due to feedback perturbation. The reduction in GMF under MEKi also provides qualitative functional support for the relevance of negative feedback in the control condition, as proposed by earlier studies \cite{Corbalan-Garcia1996, Saha2012, Umeki2025}.Here, the negative feedback helps to align the SOS and RAF distributions, leading to coordinated signal transmission. Thus, INF alone mainly reports the loss of statistical coupling, whereas GMF reveals whether RAF preserves the distributional correspondence during signal transduction. This distinction is biologically relevant because RAF transmits the SOS-dependent signal toward the MEK-ERK module, which further regulates downstream gene-expression programs and cell-state decisions \cite{Murphy2002, Uhlitz2017}. Thus, changes in SOS-RAF distributional correspondence may alter the effective signal passed to downstream MAPK activity, even when statistical coupling (correlation) is partially maintained. Therefore, GMF provides an additional layer for understanding signaling performance beyond what is captured by INF alone.

\section*{Discussion}
\label{sec3}

We introduce two complementary measures of signaling fidelity: INF, quantified by MI, and GMF, quantified by the inverse of 2-WD. INF captures how well output responses discriminate distinct input states, whereas GMF evaluates how faithfully the output preserves the distributional structure of the input. Because state discrimination and distributional correspondence describe different aspects of signal transmission, neither measure alone provides a complete description of signaling reliability. The dual-fidelity framework therefore provides a way to characterize how biological systems can regulate these two requirements.

Application of the framework to canonical gene regulatory motifs shows that network topology and biochemical parameters constrain the interplay between INF and GMF. C1-FFL can access high-performance regimes in both dimensions, consistent with its role as a robust signal-processing architecture \cite{Mangan2003, Alon2007}. In contrast, feedback motifs, especially NFL, tend to prioritize GMF at the cost of INF, consistent with their ability to buffer and stabilize responses \cite{Barkai1997, Stelling2004}. This provides a possible mechanistic interpretation of why particular motifs may be favored in specific biological contexts. Different signaling tasks impose different fidelity demands: decision-like responses require reliable state discrimination, whereas buffering or homeostatic responses require stable input-output correspondence. Network topology and biochemical parameters regulate the combinations of INF and GMF that each motif can realize. These results raise the possibility that a given architecture may be favored not because it maximizes one fidelity alone, but because it can realize a particular combination of state discrimination and distributional correspondence suited to a particular cellular function.

Our experimental data analysis of the TNF signaling pathway is qualitatively consistent with the theoretical prediction for negative feedback most clearly in the late phase, while also revealing that the dual-fidelity behavior is time-dependent. An INF-only interpretation would rank the two cell types differently depending on the time point and would not account for the role of A20-mediated feedback in WT cells. Moreover, the late-phase mutant behavior would seem to outperform WT, appearing counterintuitive given that WT cells retain the feedback. This apparent paradox is clarified by considering GMF. In the late phase, WT cells preserve higher GMF when A20-mediated feedback is engaged, indicating a more stable and proportionate signal-response relationship. This observation suggests that the feedback interaction differentially modulates state discrimination and distributional correspondence in the present TNF-signaling setup.
The experimental data analysis of the RAS-MAPK signaling pathway further clarifies the significance of GMF in signal transmission beyond what INF provides. An INF-only view would mainly identify the strongest transient loss of SOS-RAF correlation along the INF trajectory. However, GMF adds a complementary view by identifying when SOS and RAF show the strongest distributional correspondence along the signaling trajectory. Moreover, the reduction in maximum GMF under MEKi indicates that MEK inhibition weakens SOS-RAF distributional alignment, which provides qualitative functional support for the relevance of the negative feedback proposed in earlier studies \cite{Corbalan-Garcia1996, Saha2012, Umeki2025}. Thus, TNF and RAS-MAPK analyses indicate that input-output distributional correspondence is an important component of signaling fidelity that complements state discrimination. Together with the motif analyses, these results are consistent with the idea that signal transmission may tune these two objectives rather than maximizing either alone.

While these theoretical and experimental analyses demonstrate the utility of the dual-fidelity framework, several limitations constrain the broader biological conclusions that can currently be drawn. First, the theoretical models analyzed here are based on the Langevin formalism, which utilizes Hill-type functions to model gene regulation. Moreover, the estimations of MI and the 2-WD for these models are based on Gaussian approximations. Thus, the fidelity trends obtained from these coarse-grained models should be interpreted as qualitative and illustrative results that provide intuition into how INF and GMF vary across minimal gene-regulatory architectures. We further note that these models do not resolve all molecular details of gene regulation; nonetheless, they can capture key features like promoter saturation through the BAP $\theta$ and cooperative binding through the Hill coefficient $h$. The analysis of C1-FFL with $h=2$ shows that weak cooperativity preserves the qualitative trade-off observed in the non-cooperative model (see \textit{Supplementary Fig.~S1}). Additional regulatory layers, such as promoter-state switching, chromatin accessibility and modification, transcriptional bursting, time delays, spatial compartmentalization, pathway crosstalk, and multiple nested feedback interactions, are not represented in the present models. Because these regulatory processes can modify noise statistics, temporal dynamics, nonlinear responses, and effective interactions among signaling components, their omission may affect the quantitative INF-GMF relationships and, in some regimes, the motif-specific trends obtained from the present minimal models. Incorporating these layers will be an important direction for future study.

Second, the topology- and parameter-dependent INF-GMF relationships are obtained by optimizing the coarse-grained models. However, their interpretation as biological design principles, functional strategies, or evolutionary preferences remains a plausible hypothesis rather than a definitive biological conclusion. Validating such interpretations will require richer experimental datasets that relate network architecture, dual-fidelity behavior, and cellular function. Third, the experimental data analysis of the TNF-signaling system relies on the reconstructed distributions of input and output from the published summary statistics (see \hyperref[sec4]{Materials and methods}). The resulting INF and GMF values should therefore be interpreted as model-assisted estimates rather than direct measurements of the full distributions. Moreover, in the RAS-MAPK system, the fidelities are estimated using the Gaussian-approximated expressions of MI and the 2-WD (see \hyperref[sec4]{Materials and methods}). Thus, the experimental analyses demonstrate that GMF captures distributional features complementary to those quantified by INF. Establishing whether cells actively exploit this dimension of fidelity, and whether GMF has a causal biological role, remains an important direction for future investigation. Such validation will require experiments that directly relate GMF to measurable signaling functions and downstream cellular outcomes.

Among these limitations, the Gaussian approximation requires further consideration. While the Gaussian framework captures many biologically relevant regimes, the distributions often deviate due to gene expression bursting or multi-stability. In such cases, MI should be computed from the full joint distribution (Eq.~(\ref{eqm2})), and the 2-WD should be evaluated using its general optimal-transport formulation (Eq.~(\ref{eqm3})), typically necessitating numerical methods. Given that the channel $P(z|x)$ is Gaussian, MI is maximized by a Gaussian input distribution $P_X(x)$ for fixed covariance. Under this assumption, the Gaussian evaluated MI, $I_{\text{Gaussian}}(X;Z)$, provides an upper bound on the true INF achievable by any input distribution with the same covariance, i.e., $I_{\text{Gaussian}}(X;Z) \ge I (X;Z)$ \cite{Cover1991}. The Gaussian-approximated 2-WD admits a lower bound to the true 2-WD \cite{Gelbrich1990}, i.e., $W_{\text{Gaussian}}(X,Z) \le W (X,Z)$, which gives $W_{\text{Gaussian}}(X,Z)^{-1} \ge W(X,Z)^{-1}$. Combining these two bounds, the Gaussian evaluated Lagrangian $\mathcal{L}_{\rm Gaussian} := I_{\text{Gaussian}}(X;Z) - \lambda [W_{\text{Gaussian}}(X,Z)]^2$ provides an upper bound on the true objective $\mathcal{L}$ for fixed means and covariances. Therefore, our Gaussian analysis yields an upper bound on the achievable dual-fidelity, and the qualitative conclusions drawn from it remain valid. When $P(z|x)$ deviates substantially from a Gaussian form, however, no general upper bound on $I(X;Z)$ and hence on $\mathcal{L}$ is guaranteed, motivating numerical evaluation using full distributions. Such non-Gaussian regimes are biologically important because they can reflect bursting, multistability, switch-like activation, or heterogeneous cell states. We therefore add an illustrative example in the \textit{Supplementary Sec.~S6}, showing that two systems can have comparable MI but markedly different 2-WD when the same unimodal input generates either a unimodal or bimodal output. Thus, GMF captures response redistribution beyond what is resolved by MI alone, and thereby links distributional correspondence to signaling outcomes.

Having defined the scope of the present results, we next consider the broader theoretical and practical advantages of the dual-fidelity framework. In particular, the 2-WD-based Lagrangian has a wider connection to thermodynamic trade-off inequalities and observable-dependent distortion measures.
According to the analogy of thermodynamic trade-off relations \cite{ito2024geometric,nagayama2025geometric}, the 2-WD-based Lagrangian $\mathcal{L}$ can provide a universal lower bound for various other objective functions in the rate-distortion theory (RDT). The 2-WD is related to the 1-Wasserstein distance $W_1(X,Z)$, which is given by the dual optimization problem (Kantorovich-Rubinstein duality~\cite{Villani2008}), and it is known that the inequality $W(X,Z) \geq W_1(X,Z) \geq \int d x d z (h(x) - h(z)) P_{X,Z}(x,z) =: D_h(X,Z)$ holds for any observable $h$ satisfying the 1-Lipschitz condition $\|\nabla h\| \leq 1$. Therefore, the inequality $\mathcal{L}_{h} \geq \mathcal{L}$ always holds for the objective function $\mathcal{L}_{h}:=I(X;Z)- \lambda [D_h(X,Z)]^2$ using other distortions $[D_h(X,Z)]^2$. This implies that the behavior can be similar to that observed with $\mathcal{L}$ even when using a different objective function $\mathcal{L}_{h}$ to describe changes in distribution. This suggests that our results could be robust to differences in various definitions for the distribution changes. This connection also clarifies how our formulation differs from classical RDT, where the distortion function is usually specified a priori for the particular system. The objective in RDT is to minimize the information required to achieve an allowed distortion. In contrast, our framework tunes both MI and the 2-WD. Because the 2-WD bounds a class of observable-dependent distortions $D_h(X,Z)$, it provides a geometric way to constrain smooth distributional changes without choosing a separate distortion function for each observable.

Beyond this connection to rate-distortion theory, our framework is conceptually related to the work of Kolchinsky and Corominas-Murtra \cite{Kolchinsky2020}, which decomposed MI into copied and transformed components. Their approach asks whether the information transmitted from source to destination preserves the identity or similarity structure of the source messages. In sharp contrast, our framework retains MI as a measure of state discrimination and introduces the 2-WD as an additional measure of input-output distributional mismatch. 

From a practical perspective, the dual-fidelity framework also suggests future directions for experimental analysis and synthetic design. Because GMF depends on marginal distributions, it can be estimated from fluorescence intensity data in single-cell or population-level experiments. INF and GMF estimation can reveal whether a system prioritizes state discrimination, distributional correspondence, or different combinations of the two. For synthetic circuits, the same framework suggests how circuit parameters can be tuned according to task: decision-like circuits may favor high INF, whereas circuits designed to preserve input distributional features may favor high GMF. Thus, the framework provides an experimentally accessible guide for analyzing natural networks and engineering synthetic circuits.

In summary, our study indicates that distinct signaling behaviors are not fully characterized by the correlations that increase MI. By introducing GMF through the 2-WD, this work identifies input-output distributional correspondence as a distinct dimension of signaling fidelity and places OT theory as a complementary quantitative framework alongside information theory for studying cell signaling.

\phantomsection
\section*{Materials and methods}
\label{sec4}

\noindent \textbf{Modeling genetic interactions.} The stochastic dynamics of each network are governed by the Langevin equation shown in Fig.~\ref{f2}h, with the copy number vector $\bm{n}=(y,z)^\top$ and degradation rate vector $\bm{g}=(g_Y,g_Z)^\top$, where $\top$ denotes the vector transpose (motif-specific forms are tabulated in \textit{Supplementary Table~S1}). Although the dynamical form remains the same across networks, the regulatory structure is encoded in the production rate vector $\bm{f}=(f_Y,f_Z)^\top$, which varies with topology. $f_M$ ($M\in \{ Y,Z \}$) is modeled using Hill-type functions for both activation and repression (see \textit{Supplementary Table~S1}), and transcription factor binding dynamics are coarse-grained into its effective nonlinearities for analytical tractability. The steady-state mean levels are denoted by $\mu_M$. All copy numbers are dimensionless and represent absolute molecule counts. 

To account for intrinsic molecular fluctuations, we adopt a stochastic framework for $Y$ and $Z$, and the input $X$ is modeled as a Gaussian-distributed non-dynamical (static) random variable, $x \sim \mathcal{N}(\mu_X,\sigma_X^2)$, with mean $\mu_X$ and variance $\sigma_X^2$. Although $X$ is static within a single cell, its cell-to-cell variation propagates as extrinsic noise to $Y$ and $Z$. The dynamics of $Y$ and $Z$ follow Langevin-type differential equations (see Fig.~\ref{f2}h) of the form \cite{Bintu2005, Ziv2007, Tkacik2008a, Tkacik2008b, deRonde2012, Walczak2012},
\begin{equation}
    \frac{d\bm{n}}{dt}=\bm{f}-\bm{g} + \sqrt{2} \mathbf{\Gamma} \bm{\xi}(t),
    \label{eqm5}
\end{equation}
where $\bm{\xi}(t)=(\xi_Y(t),\xi_Z(t))^\top$ denotes a Gaussian white noise vector satisfying $\overline{\xi_M(t)}=0$ and $\overline{\xi_M(t)\xi_{M'}(t^\prime)}=\delta_{MM'}\delta(t-t^\prime)$ with $M\in \{Y,Z\}$ and $M'\in \{Y,Z\}$. Here, the overbar $\overline{\cdots}$ is used to denote the expectation value. The statistical properties imply that the noise processes are temporally uncorrelated and independent across species \cite{Elf2003, Paulsson2004, Swain2004, Tanase2006, Kampen2007, deRonde2010}. We write the stochastic term as $\sqrt{2} \mathbf{\Gamma} \bm{\xi}(t)$, where $\mathbf{\Gamma} = \mathrm{diag}(\Gamma_Y,\Gamma_Z)$ is a noise-amplitude matrix. With this convention, the corresponding diffusion matrix in the associated Fokker-Planck description is $\mathbf{D}=\mathbf{\Gamma}\mathbf{\Gamma}^\top$. Due to LNA, $\mathbf{D}=\mathrm{diag}(D_Y,D_Z)$ at steady-state is given by $D_Y = [f_Y(\mu_X,\mu_Z) + g_Y(\mu_Y)]/2$ and $D_Z = [f_Z(\mu_X,\mu_Y) + g_Z(\mu_Z)]/2$ \cite{Swain2004}, where $f_Y(\mu_X,\mu_Z)$ and $f_Z(\mu_X,\mu_Y)$ denote the production rates of Y and Z, respectively, and $g_Y(\mu_Y)$ and $g_Z(\mu_Z)$ represent corresponding degradation rates, evaluated in the steady-state (see \textit{Supplementary Table~S1}).

We apply the LNA~\cite{Kampen2007, Gardiner2009} to Eq.~(\ref{eqm5}) to derive analytical expressions for the statistical moments associated with each variable in the steady-state (see \textit{Supplementary Sec.~S1}). This approach is valid in the mesoscopic regime (molecule count $\sim$ hundreds) \cite{Tanase2006, Mehta2009}, where the Langevin framework approximates the chemical master equation and yields the covariance matrix upon LNA \cite{Elf2003, Paulsson2004, Kampen2007, Gardiner2009, Walczak2012}. The noise in copy numbers of $L$ is introduced as $\eta_L^2=\sigma_L^2/\mu_L^2$ and $\zeta_{LL'}=\sigma_{LL'}/(\mu_L\mu_{L'})$ ($L\in\{X,Y,Z\}$ and $L'\in\{X,Y,Z\}$), where $\sigma_L^2$ and $\sigma_{LL'}$ denote the variance and covariance, and $\mu_L$ corresponds the mean value in the steady-state. Here, the noise intensity measured by $\eta_L^2$ is defined by the squared coefficient of variation (CV$^2$), and $\zeta_{LL'}$ represents the normalized covariance. These quantities form the basis of our analytical calculations of MI and the 2-WD.

\noindent \textbf{MI and the 2-WD under Gaussian approximation.} Under Gaussian channel approximation with Gaussian input, MI can be analytically characterized using properties of multivariate normal distribution as \cite{Nandi2024},
\begin{equation}
    I(X;Z)=\frac{1}{2} \log_2 \left( \frac{\eta_X^2 \eta_Z^2}{\eta_X^2 \eta_Z^2 - \zeta_{XZ}^2}
    \right).
    \label{eqm6}
\end{equation}

\noindent We note that under the Gaussian approximation, MI is generally defined in terms of variances and covariances \cite{Cover1991, Tostevin2010}; we express it here in terms of CV$^2$ to explicitly connect correlation-dependent changes in MI with molecular noise, thereby providing a noise-centric representation.

The 2-WD between marginally Gaussian $X$ and $Z$ is given by \cite{Gelbrich1990},
\begin{eqnarray}
    W(X,Z)= \sqrt{(\mu_X-\mu_Z)^2 + (\sigma_X-\sigma_Z)^2}.
    \label{eqm7}
\end{eqnarray}

\noindent Substituting $\sigma_L = \mu_L \eta_L$, this becomes,
\begin{eqnarray}
    W(X,Z) = \sqrt{ \mathcal{V}_X + \mathcal{V}_Z - 2 \mathcal{C}_{XZ}},
    \label{eqm8}
\end{eqnarray}

\noindent where, $\mathcal{V}_X=(1+\eta_X^2)\mu_X^2$, $\mathcal{V}_Z=(1+\eta_Z^2)\mu_Z^2$, and $\mathcal{C}_{XZ}=\left( 1+\sqrt{\eta_X^2 \eta_Z^2} \right) \mu_X \mu_Z$. These closed-form expressions, Eqs.~(\ref{eqm6}) and (\ref{eqm8}), enable us to evaluate MI and the 2-WD directly from steady-state means and noise, allowing a quantitative comparison of INF and GMF across network architectures. The Gaussian channel approximation is valid in the steady-state of the network motifs, where copy numbers are kept on the order of hundreds in this study, and the LNA predicts Gaussian fluctuations. Such an approach provides a principled route to analytically quantify how each regulatory motif balances INF and GMF under biochemical noise.

With the help of these closed-form expressions, we emphasize that the trade-off between INF and GMF, observed in Fig.~\ref{f4}, is not a trivial consequence of the Lagrangian defined in Eq.~(\ref{eqm1}). To see this explicitly, we consider a limiting case where the input and output marginals are identical through equal mean $\mu_X=\mu_Z=\mu$ and equal standard deviation $\sigma_X=\sigma_Z=\sigma$. Equivalently, since $\sigma_L = \mu_L \eta_L$, the noise amplitudes become $\eta_X^2=\eta_Z^2=\eta^2$. Under these conditions, 2-WD vanishes, i.e., $W(X,Z)=0$ according to Eq.~(\ref{eqm7}). However, the MI in Eq.~(\ref{eqm6}) becomes,
\begin{equation*}
    I(X;Z) = \frac{1}{2} \log_2 \left( \frac{\eta^4}{\eta^4 - \zeta_{XZ}^2} \right)=-\frac{1}{2} \log_2 \left( 1-\rho^2 \right),
\end{equation*}

\noindent where $\rho=\zeta_{XZ}/(\eta_X\eta_Z)=\zeta_{XZ}/\eta^2$ is the correlation coefficient. Therefore, even when the input and output marginals are identical and $W(X,Z)=0$, the MI can vary through the covariance structure of the joint distribution. We note that the GMF diverges when $W(X,Z)=0$, which should be interpreted as the limiting case of maximal GMF. This example shows that the Lagrangian does not implicitly impose a trade-off between the two fidelities. Instead, the trade-off observed in the motif analysis arises from the specific biophysical and topological constraints that determine how MI and 2-WD co-vary in each model.

\noindent \textbf{Processing and analysis of TNF signaling data.} We analyze single-cell data of the TNF––NF-$\kappa$B (TNF--ATF-2) signaling network, measured in thousands of individual mouse fibroblast cells for WT and A20$^{-/-}$ cell types (Fig.~\ref{f5}a), reported in Cheong \textit{et al.} \cite{Cheong2011}. The analysis is performed at 30 minutes and 4 hours after TNF stimulation to distinguish the early response regime, where A20-mediated feedback is weak or only partially established, from the late response regime, where A20-mediated negative feedback is strongly engaged.

We use the published dose-response curves and response statistics to construct the profiles of mean concentration and CV of NF-$\kappa$B and ATF-2 for both WT and A20$^{-/-}$ cells (Fig.~\ref{f5}c,e). The CV is defined as the ratio of the standard deviation to the mean. See \textit{Supplementary Sec.~S7} for the profiles of response statistics, extracted from Cheong \textit{et al.} We note that in the original experiment, NF-$\kappa$B and ATF-2 responses were measured as fluorescence intensities in arbitrary units (a.u.), which can serve as proxies for molecular concentrations. To be specific, for each TNF dose, the experiment recorded these fluorescence intensities from thousands of cells and yielded the concentration distribution of the outputs. The mean concentration, therefore, represents the average fluorescence intensity across the cell population, and the standard deviation captures how much individual cells deviate from that mean. These experimental values, therefore, represent ensemble statistics of the cell population. 

Since the study did not report the empirical single-cell distributions, we use the reported mean concentrations and standard deviations of NF-$\kappa$B and ATF-2 to construct the output distribution at each TNF concentration, $P_{Z|X}(z|x)$, as a Gaussian. To obtain the input and output marginals, $P_X(x)$ and $P_Z(z)$, we apply an MI-constrained reconstruction strategy, where the input marginal is optimized until the MI calculated from our approximate distributions matches the MI values reported in Cheong \textit{et al.} (see \textit{Supplementary Sec.~S7}). This procedure yields joint and marginal distributions that are consistent with both the published statistics and the reported MI. From these reconstructed marginals, we then compute the 2-WD using the quantile-based definition, which is appropriate here since both input and output distributions are one-dimensional. We note that the TNF concentrations ($X$) are expressed in ng/mL, while response levels ($Z$) are measured in arbitrary units (a.u.). We, thus, rescale the response quantiles to the input scale before computing the 2-WD (see \textit{Supplementary Sec.~S7}). It is important to note that this procedure is distinct from the variational optimization used in the theoretical analysis. For the TNF-signaling data, the reported dose-response statistics define fixed condition- and time-specific distributions, rather than a tunable parameterized model. Therefore, the Lagrangian is not maximized, and neither the input noise intensity, the parameter $\lambda$, nor any biochemical parameter is varied. MI and 2-WD are instead evaluated for each experimentally defined condition and time point and should therefore be interpreted as empirical dual-fidelity diagnostics rather than optimized fidelities.

\noindent \textbf{Processing and analysis of RAS-MAPK signaling data.} We analyze the published single-cell data of EGF-stimulated SOS and RAF membrane translocation in the RAS-MAPK signaling pathway under control and MEKi conditions, reported by Umeki \textit{et al.} \cite{Umeki2025}. From the published time-series data, for each condition, paired SOS and RAF translocation values are extracted across the single-cell ensemble at each measured time point. We assign SOS translocation as the input variable $X$, and RAF translocation as the output variable $Z$. For each time point $t$ and experimental condition, the paired ensemble of SOS and RAF is used to compute the mean translocation levels $\mu_X(t)$ and $\mu_Z(t)$, the corresponding standard deviations $\sigma_X(t)$ and $\sigma_Z(t)$, and the Pearson correlation coefficient $\rho_{XZ}(t)$ between SOS and RAF. Here, the correlation coefficient is defined as $\rho_{XZ}=\sigma_{XZ}/(\sigma_X \sigma_Z)$, where $\sigma_{XZ}$ refers to the covariance between SOS and RAF. The extracted time-series profiles, together with representative joint and marginal distributions of SOS and RAF translocation, are shown in \textit{Supplementary Fig.~S6}. 

The empirical joint and marginal distributions are found to be reasonably approximated by Gaussian distributions. Therefore, for each condition and time point, the joint distribution $P_{XZ}(x,z)$ is approximated as a bivariate Gaussian specified by $\mu_X$, $\mu_Z$, $\sigma_X$, $\sigma_Z$, and $\rho_{XZ}$. INF is then computed from the Gaussian MI expression $I(X;Z) = - (1/2) \log_2 \left( 1-\rho_{XZ}^2 \right)$, which can easily be derived from Eq.~(\ref{eqm6}) by substituting $\rho_{XZ}^2 := \zeta_{XZ}^2/(\eta_X^2 \eta_Z^2) \equiv \sigma_{XZ}^2/(\sigma_X^2 \sigma_Z^2)$ where $\eta_L=\sigma_L/\mu_L$ [$L\in \{X,Z\}$] and $\zeta_{XZ}=\sigma_{XZ}/(\mu_X \mu_Z)$. GMF is computed from the inverse 2-WD between the one-dimensional Gaussian marginals of SOS and RAF, using Eq.~(\ref{eqm7}). Because SOS and RAF translocation values were reported on the same normalized fluorescence scale \cite{Umeki2025}, no additional rescaling is applied before computing the 2-WD. 

This analysis therefore provides time-resolved empirical estimates of INF and GMF directly from the measured SOS-RAF single-cell response ensemble. Similar to the TNF data analysis, no Lagrangian optimization has been performed, and no model parameter is varied in this RAS-MAPK analysis. The computed fidelities are used only as empirical diagnostics of how the SOS-RAF relationship changes over time under the two experimentally measured conditions. The error bars in both INF and GMF are estimated using bootstrap resampling.



\section*{Acknowledgements}
S.I. is supported by JSPS KAKENHI Grants
No. 22H01141, No. 23H00467, and No. 24H00834, and
UTEC-UTokyo FSI Research Grant Program, and JST
ERATO Grant Number JPMJER2302. 

\section*{Author contributions}
MN and SI designed the research and wrote the paper. MN performed the theoretical analysis and analyzed data.

\section*{Competing interests}
The authors declare no competing interests.

\end{bibunit}

\newpage

\setcounter{section}{0}
\setcounter{figure}{0}
\setcounter{table}{0}
\setcounter{equation}{0}

\begin{bibunit}

\renewcommand{\thesection}{S\arabic{section}} 
\renewcommand{\thetable}{S\arabic{table}}  
\renewcommand{\thefigure}{S\arabic{figure}} 
\renewcommand{\theequation}{S\arabic{equation}}

\section*{{\Large Supplementary information for ``Decoding cell signaling via optimal transport and information theory''}}

\noindent \textsf{\textbf{Mintu Nandi and Sosuke Ito}}

\vspace{20.0 pt}


\section{Analytical formulation of steady-state noise}

In this study, we employ linear noise approximation (LNA) \cite{Kampen2007, Gardiner2009} on the Langevin equation (see Fig.~3h and Eq.~(7) in the main text) to derive analytical expressions for the steady-state fluctuations (noise) of the network motifs. We adopt the following notations: $M \in \{Y,Z\}$ as the indices for dynamical variables, $L \in \{X,Y,Z\}$ as the indices for all variables, $n_M \in \{y,z\}$ as the corresponding state space of $M$ in copy-number units, and $n_L \in \{x,y,z\}$ as the corresponding state space of $L$ in copy-number units.

We treat the input $X$ of the network motifs as a static source of extrinsic noise. This extrinsic noise is modeled by drawing a random variable $x$ from a Gaussian distribution with the mean $\mu_X$ and variance $\sigma_X^2$, i.e., $x \sim \mathcal{N}(\mu_X,\sigma_X^2)$. We note that the value of $x$ is fixed in a single cell. However, individual cells can experience different values of $x$ following this distribution. Such an assumption makes the input $X$ an extrinsically varying random variable to the downstream stochastic components $Y$ and $Z$ in the network motifs.

The dynamics of the downstream nodes $Y$ and $Z$ are modeled using the Langevin equations (see Fig.~3h and Eq.~(7) in the main text), given by
\begin{eqnarray}
    \frac{dy}{dt} &=& f_Y(x,z) - g_Y(y) + \sqrt{2} \Gamma_Y \xi_Y(t),
    \label{eq1} \\
    \frac{dz}{dt} &=& f_Z(x,y) - g_Z(z) + \sqrt{2} \Gamma_Z \xi_Z(t),
    \label{eq2}
\end{eqnarray}

\noindent where, $f_M(\cdot)$ and $g_M(\cdot)$ represent the production and degradation rates, respectively (see Table~\ref{t1}). Here, $y$ and $z$ represent the copy number of the stochastic variables $Y$ and $Z$, respectively, with steady-state mean $\mu_M$.  The noise terms $\xi_Y(t)$ and $\xi_Z(t)$ are independent Gaussian white noise processes with zero mean and unit variance, satisfying $\overline{\xi_M(t)\xi_{M'}(t')}=\delta_{MM'}\delta(t-t')$, where $M^\prime \in \{ Y,Z \}$ \cite{Elf2003, Paulsson2004, Tanase2006, Kampen2007, deRonde2010}. Here, $\Gamma_Y$ and $\Gamma_Z$ characterize the strength of the intrinsic stochasticity in the dynamics of $Y$ and $Z$, respectively.

We note that this framework approximates the underlying master equation by treating the system as a continuous Markov process driven by Gaussian white noise. The usual derivation of the LNA proceeds with the system size expansion of the underlying master equation, which yields the Fokker-Planck equation for the stochastic fluctuations \cite{Kampen2007, Gardiner2009}. From this Fokker-Planck equation, one can write the steady-state Lyapunov equation for the covariance matrix, which, upon solving, provides the closed-form expressions of the steady-state variances and covariances in gene regulatory motifs (the derivation is scripted in \cite{Nandi2024}). In the present study, we adopt a simplified approach by applying the LNA to the Langevin equations [Eqs.~(\ref{eq1})~and~(\ref{eq2})] to derive the steady-state variances and covariances of the network components. This simplified method approximates the steady-state statistics well under the assumption that the copy numbers of the stochastic variables are sufficiently large and the fluctuations around the mean are small, as described in earlier studies \cite{Elf2003, Wallace2010, Wallace2012, Mugler2013}.

To apply the LNA, we define the stochastic fluctuations in $Y$ and $Z$ around their mean by $\delta y=y-\mu_Y$ and $\delta z=z-\mu_Z$, under the assumption that $\delta y$ and $\delta z$ are sufficiently small. We note that in the standard LNA formalism, the copy number of each species is decomposed as $n_M = \Omega \phi_M + \Omega^{1/2} \epsilon_M$, where $\phi_M$ denotes the macroscopic concentration, $\epsilon_M$ represents the mesoscopic fluctuation, and $\Omega$ refers to the system size. Now defining $\mu_M:=\Omega \phi_M$ and $\delta n_M:=\Omega^{1/2} \epsilon_M$, we obtain the relative fluctuation scales as $\delta n_M / \mu_M = \mathcal{O}(\Omega^{-1/2})$, which verifies our definition of fluctuations in the limit of large $\Omega$ with the standard LNA formalism \cite{Kampen2007, Gardiner2009}. 
Additionally, we also define the static extrinsic fluctuation of input $X$ by $\delta x = x - \mu_X$, with the assumption that $\delta x$ is small enough so that linear response in $\delta x$ remains valid for large $\mu_X$. We note here that $\delta x$ propagates to the downstream nodes $Y$ and $Z$ through the production rate function $f_M(\cdot)$. 

Now, by linearizing the production [$f_M(\cdot)$] and degradation [$g_M(\cdot)$] functions up to first order using Taylor expansion around the steady-states, we obtain,
\begin{eqnarray}
    f_Y(x,z) &=& f_Y(\mu_X+\delta x, \mu_Z+\delta z) = 
    f_Y(\mu_X,\mu_Z) + f^\prime_{YX}(\mu_X,\mu_Z) \delta x + f^\prime_{YZ}(\mu_X,\mu_Z) \delta z
    \label{eq3} \\
    f_Z(x,y) &=& f_Z(\mu_X+\delta x, \mu_Y+\delta y) = 
    f_Z(\mu_X,\mu_Y) + f^\prime_{ZX}(\mu_X,\mu_Y) \delta x + f^\prime_{ZY}(\mu_X,\mu_Y) \delta y
    \label{eq4}
\end{eqnarray}

\noindent where $f_Y(\mu_X,\mu_Z)$ and $f_Z(\mu_X,\mu_Y)$ denote the production rates of $Y$ and $Z$ at the steady-state. Furthermore, $f^\prime_{ML}$ characterizes the regulatory sensitivities and is defined as the partial derivative of $f_M$ with respect to $n_L$ evaluated in the steady-state, i.e., $f^\prime_{ML}= \partial f_M / \partial n_L$. For the degradation rates, which are first-order reactions (see Table~\ref{t1}), the expansion is exact:
\begin{eqnarray}
    g_Y(y) &=& g_Y(\mu_Y+\delta y) = \tau_Y^{-1}(\mu_Y+\delta y),
    \label{eq5} \\
    g_Z(z) &=& g_Z(\mu_Z+\delta z) = \tau_Z^{-1}(\mu_Z+\delta z),
    \label{eq6}
\end{eqnarray}

\noindent where, $\tau_M^{-1}$ is the degradation rate constant.

The expressions obtained by applying the expected value to both sides of Eqs.~(\ref{eq1})~and~(\ref{eq2}) can be approximately written as,
\begin{eqnarray}
    \frac{d \mu_Y}{dt} &=& f_Y(\mu_X,\mu_Z) - \tau_Y^{-1} \mu_Y ,
    \label{eq7} \\
    \frac{d\mu_Z}{dt} &=& f_Z(\mu_X,\mu_Y) - \tau_Z^{-1} \mu_Z ,
    \label{eq8}
\end{eqnarray}

\noindent where we used the expressions of the first-order Taylor expansions [Eqs.~(\ref{eq3} -\ref{eq6})] and $\overline{\delta x} =\overline{\delta y}=\overline{\delta z}=0$.

We, now, rewrite Eqs.~(\ref{eq1})~and~(\ref{eq2}) in terms of the fluctuations $\delta y$ and $\delta z$ by using Eqs.~(\ref{eq3}-\ref{eq8}), we obtain the linearized equations of the form,
\begin{eqnarray}
    \frac{d \delta y}{dt} &=& f^\prime_{YX}(\mu_X,\mu_Z) \delta x + f^\prime_{YZ}(\mu_X,\mu_Z) \delta z - \tau_Y^{-1} \delta y + \sqrt{2} \Gamma_Y \xi_Y(t),
    \label{eq9} \\
    \frac{d \delta z}{dt} &=& f^\prime_{ZX}(\mu_X,\mu_Y) \delta x + f^\prime_{ZY}(\mu_X,\mu_Y) \delta y - \tau_Z^{-1} \delta z + \sqrt{2} \Gamma_Z \xi_Z(t).
    \label{eq10}
\end{eqnarray}

\noindent Writing these equations in the matrix form, we obtain,
\begin{eqnarray}
    \frac{d\bm{\delta}}{dt} =  (\delta x )\bm{a} + \mathbf{J}\bm{\delta} + \sqrt{2} \mathbf{\Gamma}\bm{\xi}(t),
    \label{eq11}
\end{eqnarray}

\noindent where,
\begin{eqnarray*}
    \bm{\delta} &=& \left( 
\begin{array}{c}
    \delta y \\
    \delta z \\
\end{array} 
\right),
\quad
\bm{a} = \left( 
\begin{array}{c}
    f^\prime_{YX}(\mu_X,\mu_Z) \\
    f^\prime_{ZX}(\mu_X,\mu_Y) \\
\end{array} 
\right),
\quad
\mathbf{J} = \left( 
\begin{array}{cc}
    -\tau_Y^{-1} & f^\prime_{YZ}(\mu_X,\mu_Z) \\
    f^\prime_{ZY}(\mu_X,\mu_Y) & -\tau_Z^{-1} \\
\end{array} 
\right),
\\
\mathbf{\Gamma} &=& \left( 
\begin{array}{cc}
    \Gamma_Y & 0 \\
    0 & \Gamma_Z \\
\end{array} 
\right),
\quad
\bm{\xi}(t) = \left( 
\begin{array}{c}
    \xi_Y (t) \\
    \xi_Z (t)\\
\end{array} 
\right).
\end{eqnarray*}

\noindent Here $\mathbf{J}$ represents the Jacobian evaluated in the steady-state. To ensure the steady-state is stable, we assume $\det(\mathbf{J})=\tau_Y^{-1}\tau_Z^{-1} - f^\prime_{YZ}(\mu_X,\mu_Z)  f^\prime_{ZY}(\mu_X,\mu_Y) > 0$.


\begin{table}
\scriptsize
\renewcommand{\arraystretch}{2.5}
\caption{\textbf{Production and degradation functions for different network motifs.} 
$\alpha_M$ and $\tau_M^{-1}$ denote production and degradation rate constants, respectively, and $K_{LM}$ is the dissociation constant of regulator $L$ from the promoter of gene encoding $M$. The parameters $K_{XY}$ and $K_{ZY}$ define the BAPs $\theta_X$ and $\theta_Z$, respectively. We note that, in general, the production function can be written in the product of the Hill forms $f_M (n_{L})= \alpha_M [n_L^h/(K_{LM}^h+n_L^h)]$ for activation and $f_M (n_{L})= \alpha_M [(K_{LM}^h/(K_{LM}^h+n_L^h)]$ for repression, where the Hill coefficient $h$ captures the degree of cooperativity. In the present analysis, we use $h=1$ to ensure that the steady-state probability distributions of the components in feedback loops remain monostable.
}
\label{t1}

\begin{tabular}{lcccc}
\toprule
Function & SC, C1-FFL, I1-FFL & PFL & DNFL, NFL & \\
\midrule
$f_Y(x,z)$ 
& $\alpha_Y \frac{x}{K_{XY}+x}$ 
& $\alpha_Y \left[ \frac{x}{K_{XY}+x} + \frac{z}{K_{ZY}+z} \right]$ 
& $\alpha_Y \left[ \frac{x}{K_{XY}+x} + \frac{K_{ZY}}{K_{ZY}+z} \right]$ \\

$f_Y(\mu_X,\mu_Z)$ 
& $\alpha_Y \frac{\mu_X}{K_{XY}+\mu_X}$ 
& $\alpha_Y \left[ \frac{\mu_X}{K_{XY}+\mu_X} + \frac{\mu_Z}{K_{ZY}+\mu_Z} \right]$ 
& $\alpha_Y \left[ \frac{\mu_X}{K_{XY}+\mu_X} + \frac{K_{ZY}}{K_{ZY}+\mu_Z} \right]$ \\

$f^\prime_{YX}(\mu_X,\mu_Z)$ 
& $\alpha_Y \frac{K_{XY}}{(K_{XY}+\mu_X)^2}$ 
& $\alpha_Y \frac{K_{XY}}{(K_{XY}+\mu_X)^2}$ 
& $\alpha_Y \frac{K_{XY}}{(K_{XY}+\mu_X)^2}$ \\

$f^\prime_{YZ}(\mu_X,\mu_Z)$ 
& $0$ 
& $\alpha_Y \frac{K_{ZY}}{(K_{ZY}+\mu_Z)^2}$ 
& $-\alpha_Y \frac{K_{ZY}}{(K_{ZY}+\mu_Z)^2}$ \\
\midrule
Function & SC, PFL, NFL & DNFL & C1-FFL & I1-FFL \\
\midrule
$f_Z(x,y)$ 
& $\alpha_Z \frac{y}{K_{YZ}+y}$ 
& $\alpha_Z \frac{K_{YZ}}{K_{YZ}+y}$
& $\alpha_Z \frac{x}{K_{XZ}+x} \cdot \frac{y}{K_{YZ}+y}$ 
& $\alpha_Z \frac{x}{K_{XZ}+x} \cdot \frac{K_{YZ}}{K_{YZ}+y}$  \\

$f_Z(\mu_X,\mu_Y)$ 
& $\alpha_Z \frac{\mu_Y}{K_{YZ}+\mu_Y}$ 
& $\alpha_Z \frac{K_{YZ}}{K_{YZ}+\mu_Y}$
& $\alpha_Z \frac{\mu_X}{K_{XZ}+\mu_X} \cdot \frac{\mu_Y}{K_{YZ}+\mu_Y}$ 
& $\alpha_Z \frac{\mu_X}{K_{XZ}+\mu_X} \cdot \frac{K_{YZ}}{K_{YZ}+\mu_Y}$  \\

$f^\prime_{ZX}(\mu_X,\mu_Y)$ 
& $0$ 
& $0$ 
& $\alpha_Z \frac{K_{XZ}}{(K_{XZ}+\mu_X)^2} \cdot \frac{\mu_Y}{K_{YZ}+\mu_Y}$ 
& $\alpha_Z \frac{K_{XZ}}{(K_{XZ}+\mu_X)^2} \cdot \frac{K_{YZ}}{K_{YZ}+\mu_Y}$  \\

$f^\prime_{ZY}(\mu_X,\mu_Y)$ 
& $\alpha_Z \frac{K_{YZ}}{(K_{YZ}+\mu_Y)^2}$ 
& $-\alpha_Z \frac{K_{YZ}}{(K_{YZ}+\mu_Y)^2}$ 
& $\alpha_Z \frac{\mu_X}{K_{XZ}+\mu_X} \cdot \frac{K_{YZ}}{(K_{YZ}+\mu_Y)^2}$ 
& $-\alpha_Z \frac{\mu_X}{K_{XZ}+\mu_X} \cdot \frac{K_{YZ}}{(K_{YZ}+\mu_Y)^2}$  \\
\midrule
\multicolumn{2}{l}{Degradation functions (all motifs): \quad} 
& \multicolumn{3}{l}{$g_Y(y) = \tau_Y^{-1}y$, \quad $g_Y(\mu_Y) = \tau_Y^{-1}\mu_Y$
; \quad $g_Z(z) = \tau_Z^{-1}z$, \quad $g_Z(\mu_Z) = \tau_Z^{-1}\mu_Z$}\\
\bottomrule
\end{tabular}
\end{table}


We now recast Eq.~(\ref{eq11}) as,
\begin{eqnarray}
    d\bm{\delta} = (\delta x)\bm{a}  dt + \mathbf{J}\bm{\delta} dt + \sqrt{2} \mathbf{\Gamma} d\bm{B},
    \label{eq12}
\end{eqnarray}

\noindent where $d\bm{B}$ is the Wiener process \cite{Gillespie2000, Erban2020} that satisfies  $\overline{d\bm{B}}=\boldsymbol{0}$ and $\overline{d\bm{B}({d\bm{B}})^{\top}} = \mathbf{I} dt$. Here, $\boldsymbol{0}$ is the zero vector, and $\mathbf{I}$ is the identity matrix. Now assuming $\bm{\delta}$ to be the Ito processes, the Ito rule gives $d(\bm{\delta}\bm{\delta}^\top)=(d\bm{\delta}) \bm{\delta}^\top + \bm{\delta} ( d\bm{\delta})^\top + (d\bm{\delta}) (d\bm{\delta})^\top+ O(dt^{3/2}) = (d\bm{\delta}) \bm{\delta}^\top + \bm{\delta} ( d\bm{\delta})^\top + 2(\boldsymbol{\Gamma} d\boldsymbol{B}) (\boldsymbol{\Gamma} d\boldsymbol{B})^\top + O(dt^{3/2}) $, where $\top$ stands for the transpose of a matrix and $O(dt^{3/2})$ is Landau's O notation, which ignores orders greater than or equal to $dt^{3/2}$. Using Eq.~(\ref{eq12}) on this expansion followed by ensemble averaging yields,
\begin{eqnarray}
    \frac{d \overline{\bm{\delta}\bm{\delta}^\top}}{dt} = 
    \bm{a} \bm{\sigma}^\top + \bm{\sigma} \bm{a}^\top +
    \mathbf{J} \mathbf{\Sigma} + \mathbf{\Sigma} \mathbf{J}^\top + 2 {\mathbf{D}},
    \label{eq13}
\end{eqnarray}
where 
\begin{eqnarray*}
    \bm{\sigma} := \overline{(\delta x)\bm{\delta} } = \left( 
\begin{array}{c}
    \sigma_{XY} \\
    \sigma_{XZ} \\
\end{array} 
\right),
\end{eqnarray*}
represents a covariance vector which contains the covariances between $Y$ and $X$ and between $Z$ and $X$,
\begin{eqnarray*}
    \mathbf{\Sigma} := \overline{\bm{\delta} \bm{\delta}^\top} = \left( 
\begin{array}{cc}
    \sigma_{Y}^2 & \sigma_{YZ} \\
    \sigma_{YZ} & \sigma_{Z}^2 \\
\end{array} 
\right) ,
\end{eqnarray*}

\noindent is the covariance matrix of $Y$ and $Z$, and $\mathbf{D}:=\mathbf{\Gamma} \mathbf{\Gamma}^{\top}$ denotes the diffusion matrix that also appears in the corresponding Fokker-Planck description. 

We here calculate the variance-covariance matrix $\mathbf{\Sigma}$ and the covariance vector $\bm{\sigma}$ in the steady-state.
Due to the LNA,  ${\mathbf{D}}=\mathrm{diag}({D}_Y, {D}_Z)$ at the steady-state is given by ${D}_Y=[f_Y(\mu_X,\mu_Z)+g_Y(\mu_Y)]/2$ and ${D}_Z=[f_Z(\mu_X,\mu_Y)+g_Z(\mu_Z)]/2$ \cite{Swain2004}. Therefore, we obtain
\begin{eqnarray*}
    \mathbf{D}
= \frac{1}{2} \left( 
\begin{array}{cc}
    2\tau_Y^{-1}\mu_Y & 0 \\
    0 & 2\tau_Z^{-1}\mu_Z \\
\end{array}
\right),
\end{eqnarray*}

\noindent where we used $f_Y(\mu_X,\mu_Z) = g_Y(\mu_Y)=\tau_Y^{-1} \mu_Y$ and $f_Z(\mu_X,\mu_Y) = g_Z(\mu_Z)= \tau_Z^{-1} \mu_Z$ because $d\mu_Y/dt=d\mu_Z/dt =0$ in the steady-state. In the steady-state, $\frac{d \overline{\bm{\delta}\bm{\delta}^\top}}{dt}=0$ also gives,
\begin{eqnarray}
    \bm{a} \bm{\sigma}^\top + \bm{\sigma} \bm{a}^\top +
    \mathbf{J} \mathbf{\Sigma} + \mathbf{\Sigma} \mathbf{J}^\top + 2\mathbf{D} = 0.
    \label{eq14}
\end{eqnarray}
Solving Eq.~(\ref{eq14}) for the covariance matrix $\mathbf{\Sigma}$ yields,
\begin{eqnarray}
    \sigma_Y^2 &=& \mu_Y +
    \frac{\tau_Z^{-1} f^\prime_{YZ} \left(\mu_Y f^\prime_{ZY} + \mu_Z f^\prime_{YZ}\right)}
    {\left(\tau_Y^{-1} + \tau_Z^{-1}\right) \left(\tau_Y^{-1} \tau_Z^{-1} - f^\prime_{YZ} f^\prime_{ZY}\right)} 
    + \frac{f^\prime_{YZ} \left(\tau_Z^{-1} f^\prime_{ZX} - f^\prime_{YX} f^\prime_{ZY}\right)
    + \tau_Z^{-1} f^\prime_{YX} \left(\tau_Y^{-1} + \tau_Z^{-1}\right)}
    {\left(\tau_Y^{-1} + \tau_Z^{-1}\right) \left(\tau_Y^{-1} \tau_Z^{-1} - f^\prime_{YZ} f^\prime_{ZY}\right)}
    \sigma_{XY}
    \nonumber \\
    && + \frac{f^\prime_{YZ} \left(f^\prime_{YZ}f^\prime_{ZX} + \tau_Z^{-1} f^\prime_{YX}\right)}
    {\left(\tau_Y^{-1} + \tau_Z^{-1}\right) \left(\tau_Y^{-1} \tau_Z^{-1} - f^\prime_{YZ} f^\prime_{ZY}\right)}
    \sigma_{XZ},
    \label{eq15} \\
    \sigma_Z^2 &=& \mu_Z +
    \frac{\tau_Y^{-1} f^\prime_{ZY} \left(\mu_Y f^\prime_{ZY} + \mu_Z f^\prime_{YZ}\right)}
    {\left(\tau_Y^{-1} + \tau_Z^{-1}\right) \left(\tau_Y^{-1} \tau_Z^{-1} - f^\prime_{YZ} f^\prime_{ZY}\right)} 
    + \frac{f^\prime_{ZY} \left(f^\prime_{YX} f^\prime_{ZY} + \tau_Y^{-1} f^\prime_{ZX}\right)}
    {\left(\tau_Y^{-1} + \tau_Z^{-1}\right) \left(\tau_Y^{-1} \tau_Z^{-1} - f^\prime_{YZ} f^\prime_{ZY}\right)}
    \sigma_{XY}
    \nonumber \\
    && + \frac{f^\prime_{ZY} \left(\tau_Y^{-1} f^\prime_{YX} - f^\prime_{YZ} f^\prime_{ZX}\right)
    + \tau_Y^{-1} f^\prime_{ZX} \left(\tau_Y^{-1} + \tau_Z^{-1}\right)}
    {\left(\tau_Y^{-1} + \tau_Z^{-1}\right) \left(\tau_Y^{-1} \tau_Z^{-1} - f^\prime_{YZ} f^\prime_{ZY}\right)}
    \sigma_{XZ},
    \label{eq16} \\
    \sigma_{YZ} &=& 
    \frac{\tau_Y^{-1} \tau_Z^{-1} \left(\mu_Y f^\prime_{ZY} + \mu_Z f^\prime_{YZ}\right)}
    {\left(\tau_Y^{-1} + \tau_Z^{-1}\right) \left(\tau_Y^{-1} \tau_Z^{-1} - f^\prime_{YZ} f^\prime_{ZY}\right)} +
    \frac{\tau_Z^{-1} \left(f^\prime_{YX} f^\prime_{ZY} + \tau_Y^{-1} f^\prime_{ZX}\right)}
    {\left(\tau_Y^{-1} + \tau_Z^{-1}\right) \left(\tau_Y^{-1} \tau_Z^{-1} - f^\prime_{YZ} f^\prime_{ZY}\right)}
    \sigma_{XY}
    \nonumber \\
    && + \frac{\tau_Y^{-1} \left(f^\prime_{YZ}f^\prime_{ZX} + \tau_Z^{-1} f^\prime_{YX}\right)}
    {\left(\tau_Y^{-1} + \tau_Z^{-1}\right) \left(\tau_Y^{-1} \tau_Z^{-1} - f^\prime_{YZ} f^\prime_{ZY}\right)}
    \sigma_{XZ}.
    \label{eq17}
\end{eqnarray}

\noindent Here, we have used the following short-hand notations: $f^\prime_{YX}(\mu_X,\mu_Z) =: f^\prime_{YX}$, $f^\prime_{YZ}(\mu_X,\mu_Z)  =: f^\prime_{YZ}$, $f^\prime_{ZX}(\mu_X,\mu_Y)  =: f^\prime_{ZX}$, and $f^\prime_{ZY}(\mu_X,\mu_Y)  =: f^\prime_{ZY}$. These expressions are obtained as functions of regulatory sensitivities ($f^\prime_{ML}$), some biochemical parameters, and the elements of $\bm{\sigma}$. To derive $\bm{\sigma}$ in the steady-state, we multiply Eq.~(\ref{eq12}) by $\delta x$ followed by ensemble averaging in the steady-state, which yields,
\begin{eqnarray}
    \frac{d\bm{\sigma}}{dt}= \overline{(\delta x) \frac{d\bm{\delta}}{dt}} = 
    \bm{a} \sigma_X^2 + \mathbf{J} \bm{\sigma} =0,
    \label{eq18}
\end{eqnarray}

\noindent where $\sigma_X^2=\overline{\delta x^2}$. In Eq.~(\ref{eq18}), we use $\overline{(\delta x) \sqrt{2} \mathbf{\Gamma} \bm{\xi}}=0$ due to uncorrelation. Solving this equation yields the exact expressions of the elements of $\bm{\sigma}$ in terms of the input variance $\sigma_X^2$, as follows,
\begin{eqnarray}
    \sigma_{XY} &=& \frac{f^\prime_{YZ} f^\prime_{ZX} + \tau_Z^{-1} f^\prime_{YX}}
    {\tau_Y^{-1} \tau_Z^{-1} - f^\prime_{YZ} f^\prime_{ZY}} \sigma_X^2,
    \label{eq19} \\
    \sigma_{XZ} &=& \frac{f^\prime_{YX} f^\prime_{ZY} + \tau_Y^{-1} f^\prime_{ZX}}
    {\tau_Y^{-1} \tau_Z^{-1} - f^\prime_{YZ} f^\prime_{ZY}} \sigma_X^2.
    \label{eq20}
\end{eqnarray}

By substituting Eqs.~(\ref{eq19}) and (\ref{eq20}) into the expression for the output variance $\sigma_Z^2$, we obtain a closed-form expression of $\sigma_Z^2$ in terms of the input variance,
\begin{eqnarray}
    \sigma_Z^2 = \mu_Z + \phi_1 + \phi_2 \sigma_X^2,
    \label{eq21}
\end{eqnarray}

\noindent where,
\begin{eqnarray*}
    \phi_1 &=& \frac{\tau_Y^{-1} f^\prime_{ZY} \left(\mu_Y f^\prime_{ZY} + \mu_Z f^\prime_{YZ}\right)}
    {\left(\tau_Y^{-1} + \tau_Z^{-1}\right) \left(\tau_Y^{-1} \tau_Z^{-1} - f^\prime_{YZ} f^\prime_{ZY}\right)},
    \\
    \phi_2 &=& 
    \frac{\left(f^\prime_{YX} f^\prime_{ZY} + \tau_Y^{-1} f^\prime_{ZX}\right)^2}
    {\left(\tau_Y^{-1} \tau_Z^{-1} - f^\prime_{YZ} f^\prime_{ZY}\right)^2}.
\end{eqnarray*}

We quantify the output noise of $Z$ using the squared coefficient of variation, $\eta_Z^2=\sigma_Z^2/\mu_Z^2$. Using Eq.~(\ref{eq21}), we can write the closed-form expression of $\eta_Z^2$ as,
\begin{eqnarray}
    \eta_Z^2 = \frac{1}{\mu_Z} + \Phi_1 + \Phi_2 \eta_X^2,
    \label{eq22}
\end{eqnarray}

\noindent where, $\eta_X^2=\sigma_X^2/\mu_X^2$ characterizes the noise in the input, and the coefficients $\Phi_1=\phi_1/\mu_Z^2$ and $\Phi_2=\mu_X^2\phi_2/\mu_Z^2$ collect the contributions from the intermediate terms. The normalized covariance between $X$ and $Z$ is defined as $\zeta_{XZ}=\sigma_{XZ}/(\mu_X\mu_Z)$ and its closed-form expression can be written from Eq.~(\ref{eq20}) as,
\begin{eqnarray}
    \zeta_{XZ} = \Psi \eta_X^2,
    \label{eq23}
\end{eqnarray}

\noindent where, 
\begin{eqnarray*}
    \Psi=\frac{\mu_X \left(f^\prime_{YX} f^\prime_{ZY} + \tau_Y^{-1} f^\prime_{ZX}\right)}
    {\mu_Z \left(\tau_Y^{-1} \tau_Z^{-1} - f^\prime_{YZ} f^\prime_{ZY}\right)}.
\end{eqnarray*}

\noindent The noise of $Y$, $\eta_Y^2=\sigma_Y^2/\mu_Y^2,$ can similarly be defined from Eq.~(\ref{eq15}). $\eta_Z^2$ and $\zeta_{XZ}$ are used to calculate the mutual information (MI) and the 2-Wasserstein distance (2-WD) using Eqs.~(8)~and~(10) given in the main text.

\section{Binding affinities}

The output noise and the input–output normalized covariance depend on several biochemical parameters, including mean expression levels ($\mu_L$), the degradation rate constants ($\tau_M^{-1}$), and the regulatory sensitivities ($f^\prime_{ML}$). We refer to $f^\prime_{ML}$ as regulatory sensitivities since they capture the slope of the input-output regulatory functions in the steady-state, quantifying how strongly a small change in the regulator $L$ influences the expression of the gene $M$. These sensitivity terms themselves are functions of the production rate constants ($\alpha_M$) and the dissociation constants ($K_{LM}$) (see Table~\ref{t1}). The dissociation constant $K_{LM}$ is a coarse-grained parameter that characterizes the binding affinity of transcription factor (TF) $L$ to the promoter of the gene of $M$. A smaller $K_{LM}$ corresponds to a reduced dissociation rate, reflecting a higher affinity. In this study, we systematically vary the promoter properties of the gene encoding $Y$ (the intermediate node common to all motifs) to examine how binding affinity influences both informational and geometric fidelities.

In the simple cascade (SC), coherent type-1 feed-forward loop (C1-FFL), and incoherent type-1 feed-forward loop (I1-FFL), the promoter of $Y$ interacts with $X$, characterized by the parameter $K_{XY}$. In the positive feedback loop (PFL), double negative feedback loop (DNFL), and negative feedback loop (NFL), the promoter of $Y$ interacts with both $X$ and $Z$, governed by $K_{XY}$ and $K_{ZY}$. We tune these parameters around the half-maximal regulation ($K_{LM}=\mu_L$) \cite{Alon2006}. To quantify deviations from this reference, we define binding affinity parameters (BAPs) as $\theta_X=K_{XY}/\mu_X$ and $\theta_Z=K_{ZY}/\mu_Z$. By construction, $\theta_X=1$ and $\theta_Z=1$ corresponds to half-maximal regulation of $Y$'s promoter by $X$ and $Z$. If $\theta_X<1$ and $\theta_Z<1$, this indicates strong binding affinities of $X$ and $Z$ for the $Y$'s promoter, and if $\theta_X>1$ and $\theta_Z>1$, this represents weak binding affinities (see Fig.~4 in the main text).

The introduction of BAP as a measure of promoter affinity serves two key purposes. First, it separates binding affinity from absolute expression levels, so that comparisons across different motifs do not depend on the scale of protein copy numbers. Second, it directly links BAP to the regulatory input--output curve, where $K$ sets the half-saturation point. Functionally, promoters with strong affinities can drive rapid and decisive state changes, while weak affinities give rise to graded responses and dampened variability. Such variations can be engineered through targeted promoter mutations or the synthetic modification of TF--DNA binding sites, making it an experimentally tunable parameter.

\section{Model Parameters}
\label{secs3}

To optimize the objective function $\mathcal{L}$ with respect to the input noise $\eta_X^2$ (Eq.~(6) in the main text), we fix the mean copy number of all three proteins to $\mu_X=\mu_Y=\mu_Z=100$. This ensures that copy-number-driven noise remains consistent across species, allowing us to isolate the role of noise propagation from input to output in shaping the dual-fidelity landscape of different network motifs. Choosing copy numbers on the order of a hundred also justifies the use of Gaussian approximations \cite{Tanase2006, Mehta2009}.

The degradation rates are chosen as $\tau_Y^{-1}=1.0$ and $\tau_Z^{-1}=10.0$ to enforce a separation of timescales in which $Y$ decays slowly and $Z$ decays rapidly. Such a hierarchy is a common requirement for reliable signaling: a slower upstream component ($Y$) preserves memory of prior input ($X$), while a faster downstream component ($Z$) can rapidly track changes without accumulating excessive noise. This arrangement facilitates efficient information transmission along the motif \cite{Nandi2024}. Within this regime, we analyze how input noise can be tuned to optimize the informational and geometric fidelities. We next vary each of $\theta_X$ and $\theta_Z$ across three regimes: $<1$, $=1$, and $>1$ represented by the values 0.5, 1, and 2, respectively. We have nine possible sets of ($\theta_X,\theta_Z$) for PFL, DNFL, and NFL motifs. However, for SC, C1-FFL, and I1-FFL, we have three possible values of $\theta_X$ only. This allows us to explore how promoter affinities of $Y$ shape the balance between the two fidelities.

For each species, the production rate constants $\alpha_M$, encapsulated in ${f}_M$ (see Table~\ref{t1}), are obtained by considering Eqs.~(\ref{eq7}) and (\ref{eq8}) in the steady-state. Finally, to map the optimal dual-fidelity profiles, we vary the Lagrange multiplier $\lambda$ over the range $10^{-4}$ to $1$. When $\lambda$ is very small, the optimization is dominated by MI, and as it increases up to unity, the 2-WD progressively balances MI. We choose this range because the numerical scale of MI is much smaller than that of the 2-WD (see Fig.~5 in the main text). Restricting $\lambda$ to $10^{-4}$--$1$ ensures that both measures enter the optimization on a comparable footing despite their inherent difference in magnitude.

\begin{figure}[!t]
\centering
\includegraphics[width=0.5\columnwidth,angle=0]{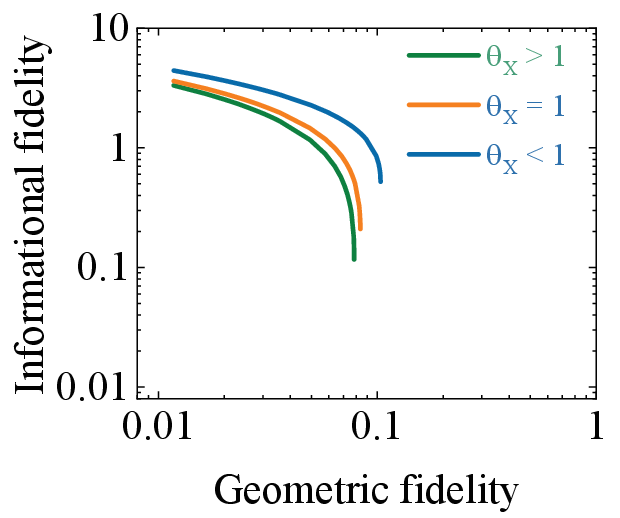}
\caption{
\textbf{Dual-fidelity behavior for the Hill coefficient $h=2$.} C1-FFL showing trade-off between informational and geometric fidelities across different BAPs. Informational fidelity is expressed in bits and geometric fidelity in $\text{(copy number)}^{-1}$.
}
\label{fs01}
\end{figure}

\section{Interpreting 2-Wasserstein distance}

The numerical value of the 2-WD is expressed in the same units as the variables being compared. In our theoretical analysis, where system components $X$ and $Z$ are expressed in molecular copy numbers, the 2-WD carries units of molecule count. In contrast, experimental estimates express the 2-WD in units of TNF concentration. The interpretation of $W(X,Z)=1$ is therefore unit-dependent, yet its conceptual meaning remains consistent: it quantifies the minimal average mismatch between the input and output distributions under an optimal transport plan.

A value of $W(X,Z)=1$ indicates that, under the optimal mapping between the input and output, the root-mean-square discrepancy between them is one unit. In copy-number units, this corresponds to an average distortion of one molecule per cell between input and output distributions. In the TNF concentration unit, it implies that the signaling pathway cannot reduce the average concentration distortion below one concentration unit, given the observed variability.

Under Gaussian approximation (see Eq.~(9) in the main text), $W(X,Z)=1$ occurs when $(\mu_X-\mu_Z)^2+(\sigma_X-\sigma_Z)^2=1$. This indicates three representative cases: (1) a pure mean shift of one unit ($\mu_Z=\mu_X \pm 1$) with equal standard deviations ($\sigma_Z=\sigma_X$), (2) a pure spread difference of one unit ($\sigma_Z=\sigma_X \pm 1$) with equal means, or (3) any combination of mean and spread differences satisfying the same relation. This clarifies the biological implications that a nonzero mean shift indicates a systematic bias in the response, while a spread difference reflects the amplification or filtering of input variability. The value $W(X,Z)=1$ thus serves as a unified measure of the total geometric distortion introduced by the signaling process, whether expressed in theoretical or experimental units.

In the theoretical analysis, we set geometric fidelity $W(X,Z)^{-1} \geq 1$ as the minimum for reliable distributional correspondence (see main text). This threshold corresponds to $W(X,Z)\le1$. Under the Gaussian form with equal means (see Sec.~\ref{sec3}), the 2-WD reduces to the absolute difference between the input and output standard deviations, so the criterion is simply $|\sigma_X - \sigma_Z| \le 1$ in the copy number unit. This allows at most a one-molecule difference in spread between the input and output. We choose this threshold because it represents a fundamental unit of discreteness in stochastic gene expression. A standard deviation distortion larger than a single molecule implies the regulatory mechanism is actively reshaping the noise profile beyond the minimal discreteness noise inherent to molecular counting. We note that the numerical value of this threshold is unit-dependent and is strictly meaningful within our theoretical framework, where the 2-WD is measured in copy-number units. In experimental settings, where variables are expressed in physical units such as ng/mL or arbitrary fluorescence units, an equivalent threshold must be defined relative to the effective resolution and fluctuation level of the measurement system, rather than by the absolute numerical value alone.

\section{Contrasting 2-Wasserstein distance with Kullback-Leibler divergence}

Here, we compare the 2-WD and Kullback-Leibler divergence (KLD) as possible measures of mismatch between the input and output marginal distributions. The purpose is to clarify why the 2-WD is used to quantify input-output distributional mismatch. 

The KLD between two marginal distributions $P_X$ and $P_Z$ can be written as
\begin{equation}
    D_{KL}(P_X\| P_Z) = \int du P_X(u) \log\left[\frac{P_X(u)}{P_Z(u)} \right],
    \label{eqs41}
\end{equation}

\noindent where both distributions are represented on a common coordinate $u$. This quantity compares the relative density mismatch. Importantly, the KLD is asymmetric: $D_{KL}(P_X\| P_Z) \neq D_{KL}(P_Z\| P_X)$ in general. Therefore, its numerical value depends on which distribution is taken as the reference. In a signaling context, this means that comparing the input to the output, $D_{KL}(P_X\| P_Z)$, and comparing the output to the input, $D_{KL}(P_Z\| P_X)$, can assign different mismatch values to the same pair of marginal distributions. In contrast, the 2-WD is defined by the optimal-transport problem (see Eq.~(3) in the main text) and, unlike the KLD, it is symmetric. Thus, 2-WD does not depend on which marginal is chosen as the reference. It measures the geometric displacement of probability mass required to transform one distribution into the other.

The distinction between the 2-WD and KLD becomes explicit in the Gaussian case, where marginals are assumed to follow a Gaussian distribution, $P_X = \mathcal{N}(\mu_X,\sigma_X^2)$ and $P_Z = \mathcal{N}(\mu_Z,\sigma_Z^2)$. In this case, the 2-WD becomes (see Eq.~(9) in the main text),
\begin{equation}
    W(X,Z) = \sqrt{\left(\mu_X-\mu_Z \right)^2 + \left(\sigma_X-\sigma_Z \right)^2}.
    \label{eqs42}
\end{equation}

\noindent Thus, the 2-WD separates the mismatch into two direct geometric contributions: displacement of the mean and mismatch in spread. By contrast, the KLD between input and output marginals becomes,
\begin{equation}
    D_{KL} = \frac{1}{2} \left( \log\left[\frac{\sigma_Z^2}{\sigma_X^2} \right] + \frac{\sigma_X^2}{\sigma_Z^2} + \frac{\left(\mu_X-\mu_Z \right)^2}{\sigma_Z^2} - 1 \right).
    \label{eqs43}
\end{equation}

\noindent Thus, the KLD decomposes directly into three interpretable contributions: a logarithmic mismatch in the variances, a ratio of variances, and a displacement of the mean scaled by variance. However, the expression in Eq.~(\ref{eqs43}) changes when the direction of comparison is reversed, that is, when the KLD is computed from the output distribution to the input distribution. The KLD, although perfectly well defined, possesses some limitations compared to 2-WD as detailed below.

To clarify the contrast between the 2-WD and KLD, we consider two limiting cases: (1) when the two marginals have equal spread $\sigma_X=\sigma_Z=\sigma$ with unequal means and (2) when the marginals have equal means $\mu_X=\mu_Z$ with unequal spread. For case-1, the 2-WD and KLD reduce to,
\begin{eqnarray}
    W(X,Z) &=& \lvert \mu_X-\mu_Z \rvert,
    \label{eqs44} \\
    D_{KL} &=& \frac{\left( \mu_X-\mu_Z \right)^2}{2\sigma^2}.
    \label{eqs45}
\end{eqnarray}

\noindent Thus, the 2-WD directly measures the physical displacement between the two distributions. However, the KLD does not directly measure displacement, instead, it measures the squared displacement normalized by the variance. Therefore, for the same physical shift $\lvert \mu_X-\mu_Z \rvert$, the KLD decreases as the distributions become broader, whereas 2-WD remains equal to the actual displacement. In the limit of very broad distributions, $\sigma \to \infty$ with fixed $\lvert \mu_X-\mu_Z \rvert$, we obtain $D_{KL} \to 0$, whereas 2-WD remains unchanged. Thus, the KLD can become arbitrarily small for a finite geometric displacement if the displacement is small relative to the distributional width. The 2-WD retains the physical magnitude of the shift.

For case-2, the 2-WD and KLD reduce to,
\begin{eqnarray}
    W(X,Z) &=& \lvert \sigma_X-\sigma_Z \rvert,
    \label{eqs46} \\
    D_{KL} &=& \frac{1}{2} \log\left[\frac{\sigma_Z^2}{\sigma_X^2} \right] + \frac{\sigma_X^2}{2\sigma_Z^2} - \frac{1}{2}.
    \label{eqs47}
\end{eqnarray}

\noindent To see the geometric implication more clearly, suppose $\sigma_Z = \epsilon \sigma_X$ with $0<\epsilon<1$. Then the 2-WD and KLD become,
\begin{eqnarray*}
    W(X,Z) &=& \lvert \sigma_X (1-\epsilon) \rvert, \\
    D_{KL} &=& \frac{1}{2} \log\left[\epsilon^2 \right] + \frac{1}{2\epsilon^2} - \frac{1}{2}.
\end{eqnarray*}

\noindent In the strong compression limit, $\epsilon \to 0$, 2-WD remains finite, but the KLD diverges. Therefore, when a broad input distribution is compressed into a very narrow output distribution, the KLD becomes very large because the output assigns extremely small probability to regions where the input still has substantial probability. In contrast, the 2-WD remains a finite geometric measure of the change in distributional spread. This distinction is relevant for signaling systems where feedback, saturation-like response compression, or regulatory constraints narrow the output distribution.

These examples show that the KLD and 2-WD quantify different aspects of marginal mismatch. The KLD is useful for measuring directed relative density mismatch, but it is not designed to measure geometric displacement between distributions. In contrast, 2-WD directly quantifies shifts, broadening, compression, and redistribution of probability mass. For this reason, we use 2-WD to quantify input-output distributional mismatch.

\section{Role of 2-Wasserstein distance beyond mutual information: an illustrative example}

The main theoretical analysis in this work uses Gaussian approximations to obtain analytically tractable expressions for MI and the 2-WD (as given in Eqs.~(8)~and~(9) in the main text). However, the dual-fidelity framework is not restricted to Gaussian signaling distributions. To illustrate how geometric fidelity can characterize signaling outcomes beyond MI in a non-Gaussian regime, we consider a simple signaling-relay example in which a unimodal upstream signal produces either a graded downstream response or a switch-like heterogeneous downstream response.

This example is not intended to model a specific biochemical pathway. Instead, it is used as a controlled calculation to show that two signaling relays can transmit comparable MI while exhibiting very different 2-WD. This directly illustrates why the 2-WD can be useful when signaling responses undergo qualitative distributional changes, such as unimodal-to-bimodal transformations.

The upstream signal $X$ is represented by a positive, unimodal, normalized activity, 
\begin{equation}
    x \sim \mathrm{Gamma}(\alpha=36,\beta=1/12),
    \label{eqs60}
\end{equation}

\noindent where $\alpha$ and $\beta$ are the shape and scale parameters, respectively. This gives the mean $\mu_X=\alpha \beta=3$ and the standard deviation $\sigma_X=\sqrt{\alpha}\beta=0.5$. This choice avoids a negative-valued input and therefore better resembles a concentration-like upstream signaling activity.

We first consider a graded signaling relay, in which the downstream response $Z_G$ follows the upstream signal continuously with additive noise,
\begin{equation}
    z_G = x + \Gamma_G \xi,
    \label{eqs61}
\end{equation}

\noindent where $\xi \sim \mathcal{N}(0,1)$ and $z_G$ denotes the state of the output $Z_G$. Equivalently, for a given input value, the conditional output can be given by 
\begin{equation}
    z_G|x \sim \mathcal{N}(x,\Gamma_G^2).
    \label{eqs62}
\end{equation}

\noindent Thus, the mean of the output response shifts continuously with the input value. The parameter $\Gamma_G$ controls the noise amplitude of the graded relay. It is adjusted numerically so that the graded relay had comparable MI to the switch-like relay described below.

\begin{figure}[!t]
\centering
\includegraphics[width=0.8\columnwidth,angle=0]{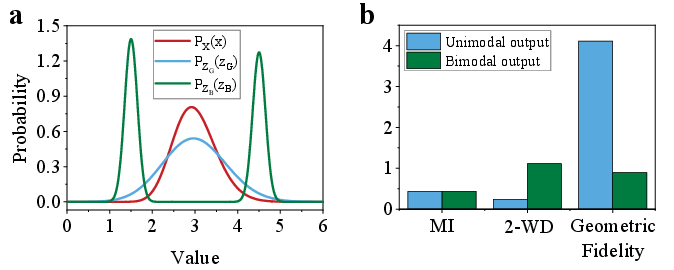}
\caption{
\textbf{Illustrative non-Gaussian example showing the added value of geometric fidelity.}
\textbf{a} Probability density profiles of the input $P_X(x)$, the graded output $P_{Z_G}(z_G)$, and the switch-like output $P_{Z_B}(z_B)$. 
\textbf{b} Comparison of MI, 2-WD, and geometric fidelity for the two relays. The colored bars show corresponding calculated values of MI using Eq.~(\ref{eqs65}), 2-WD using Eq.~(\ref{eqs66}), and GMF. Although both relays show comparable MI, the switch-like relay exhibits a substantially larger 2-WD and lower geometric fidelity. This illustrates that MI captures input-output dependence, whereas 2-WD captures geometric reorganization of the response distribution. Here, MI is expressed in bits, whereas 2-Wd and GMF are dimensionless on the normalized activity scale.
}
\label{fs001}
\end{figure}

We then consider a switch-like heterogeneous relay. In this case, the upstream signal does not directly set the mean of the downstream response. Instead, it controls the probability that the system occupies either a low-response or high-response state. To define this process, we introduce a binary response state $S\in \{0,1\}$, where $S=0$ denotes the low-response state and $S=1$ denotes the high-response state. For a given input value, the probability of occupying the high-response state is modeled by a Hill-type function,
\begin{equation}
    P(S=1|x) = s(x) = \frac{x^h}{K^h+x^h}.
    \label{eqs63}
\end{equation}

\noindent Therefore $P(S=0|x) = 1 - s(x)$. Here, we use $h=16$ and $K=3$. The Hill form is chosen because it provides a standard biochemical representation of threshold-like or cooperative activation. The parameter $K$ sets the activation threshold, and $h$ controls the sharpness of switching between the low- and high-response states. After the response state $S$ is selected, the downstream response $Z_B$ is generated from a Gaussian distribution centered around the corresponding response level. Specifically, $z_B|S=0 \sim \mathcal{N}(z_{B_{\rm low}},\sigma_B^2)$ and $z_B|S=1 \sim \mathcal{N}(z_{B_{\rm high}},\sigma_B^2)$. Here, we use $z_{B_{\rm low}}=1.5$, $z_{B_{\rm high}}=4.5$, and $\sigma_B=0.15$. Thus, for a fixed input value, $Z_B$ is generated from the low-response state with probability $1-s(x)$, and from the high-response state with probability $s(x)$. In this sense, $X$ controls the mixture weights of the low- and high-response states, rather than directly shifting the response mean as in the graded relay. Equivalently, the conditional distribution of $Z_B$ for a given input state is a two-component Gaussian mixture, given by,
\begin{equation}
    z_B|x
    \sim
    [1-s(x)]\mathcal{N}\left(z_{B_{\rm low}},\sigma_B^2\right)
    +
    s(x)\mathcal{N}\left(z_{B_{\rm high}},\sigma_B^2\right).
    \label{eqs64}
\end{equation}

\noindent This means that the response is sampled from one of two Gaussian response states, with $X$-dependent probabilities. Therefore, the same unimodal input distribution can generate a bimodal output distribution when the downstream response becomes heterogeneous. Biologically, this type of distributional transformation may correspond to threshold activation, multistability, population splitting, or heterogeneous cell-state responses.

The input and output distributions are computed numerically as follows. First, the input probability $P_X(x)$ is evaluated on a fixed grid of $x$ values following Eq.~(\ref{eqs60}). For each value of $x$, the conditional output probability $P(z|x)$, where $z \equiv z_G ~{\rm or}~ z_B$, is evaluated on a fixed grid of $z$ values following Eq.~(\ref{eqs62}) for the graded relay and Eq.~(\ref{eqs64}) for the switch-like relay. The output marginal distribution is then obtained from $P_Z(z)=\int dx P_X(x) P(z|x)$. The plots of the marginal distributions for the two systems are shown in Fig.~\ref{fs001}a.

For both relays, MI is computed from the full joint distribution. For a given conditional output distribution $P(z|x)$, where $z \equiv z_G ~{\rm or}~ z_B$, MI is defined as,
\begin{equation}
    I(X;Z) = \int dx dz P_X(x) P(z|x) \log_2 \left[\frac{P(z|x)}{P_Z(z)} \right].
    \label{eqs65}
\end{equation}

\noindent The 2-WD is computed from the one-dimensional marginal distributions using the quantile representation \cite{Peyre2019},
\begin{equation}
    W(X,Z)^2
    =
    \int_0^1
    \left[
    F_X^{-1}(u)-F_Z^{-1}(u)
    \right]^2du,
    \label{eqs66}
\end{equation}

\noindent where $F_X^{-1}$ and $F_Z^{-1}$ denote the inverse cumulative distribution functions of the input and output marginals, respectively. We note that since both variables are normalized activity variables on the same dimensionless scale, no additional rescaling was applied before computing the 2-WD.

With the chosen parameters, the switch-like heterogeneous relay gives $I(X;Z_B)=0.437$ bits and $W(X,Z_B)=1.114$. The corresponding geometric fidelity becomes $W(X,Z_B)^{-1}=0.897$. The noise amplitude of the graded relay is then tuned to $\Gamma_G=0.547$ which gives comparable MI of $I(X;Z_G)=0.437$ bits, but  $W(X,Z_G)=0.243$ and corresponding $W(X,Z_G)^{-1}=4.113$. Therefore, although the two relays transmit nearly the same amount of information about the upstream signal, the switch-like relay shows a much larger 2-WD because the downstream response distribution is reorganized into two separated activation states. The values of MI, 2-WD and geometric fidelity are compared between the two relays in Fig.~\ref{fs001}b.

This illustrative example demonstrates the distinct roles of the two fidelity measures. MI determines whether the downstream response remains statistically dependent on the upstream signal. However, it does not directly quantify how the response distribution is geometrically reorganized. In contrast, the 2-WD captures the redistribution of probability mass from a unimodal upstream signal into separated downstream response states. Thus, geometric fidelity can quantify distributional distortions such as shifts, broadening, compression, and unimodal-to-bimodal transitions. These features are relevant for biological processes involving heterogeneous activation, population splitting, or cell-state transitions.

This example therefore supports the main motivation of the dual-fidelity framework: two signaling systems can carry comparable information while differing substantially in how faithfully the downstream response preserves the distributional structure of the upstream signal. MI and the 2-WD are therefore complementary rather than redundant measures of signaling fidelity.

\begin{figure}[!t]
\centering
\includegraphics[width=1\columnwidth,angle=0]{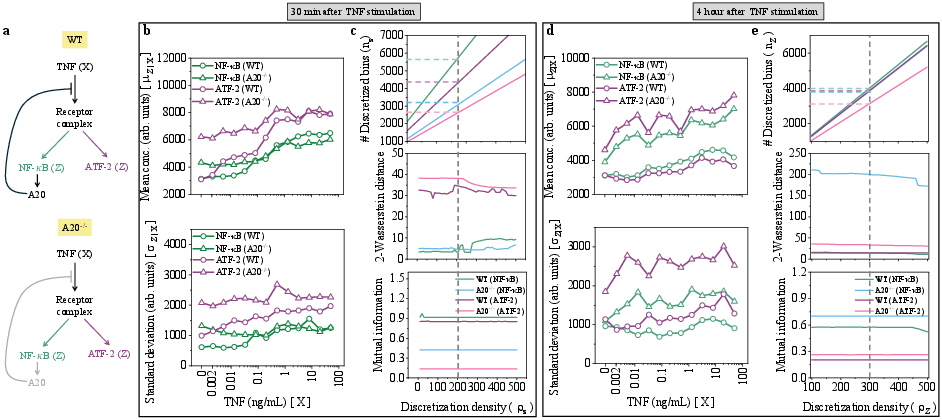}
\caption{\textbf{Experimental system and data processing.}
\textbf{a} Schematics of TNF signaling in WT and A20$^{-/-}$ cells. here, green label denotes NF-$\kappa$B and purple represents ATF-2. Moreover, the grey arrow in A20$^{-/-}$ cells indicates the absence of negative feedback.
\textbf{b} The dose-response statistics at 30 minutes after TNF stimulation.
\textbf{c} Convergence of MI and the 2-WD as a function of discretization at 30-minute time point.
\textbf{d} The dose-response statistics at 4 hours after TNF stimulation.
\textbf{e} Convergence of MI and the 2-WD as a function of discretization at 4-hour time point.
In panels \textbf{b} and \textbf{d}, open circles and open triangles denote WT and A20$^{-/-}$ cells, respectively. Each profile contains $n=13$ TNF dose values. The upper and lower plots show the published conditional mean and standard deviations, respectively. 
In panels \textbf{c} and \textbf{e}, colors identify the pathway and cell type as labeled. The vertical grey dashed lines mark the selected discretization densities $\rho_z=200$ and $300$. The horizontal colored dashed lines denote the corresponding $n_z$ values.
}
\label{fs1}
\end{figure}

\begin{figure}[!t]
\centering
\includegraphics[width=1\columnwidth,angle=0]{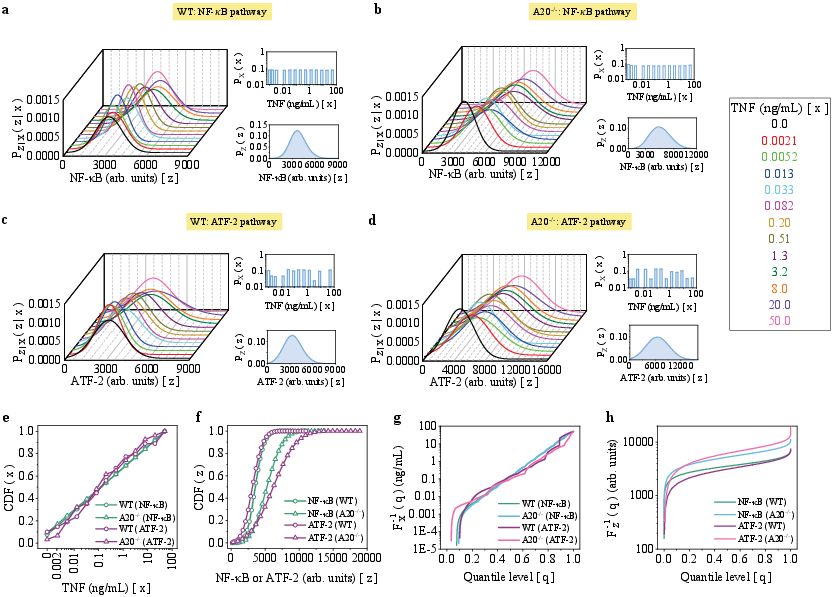}
\caption{\textbf{Representative reconstructed distributions at 4-hour time point.}
Gaussian conditionals, optimized input distributions, and output marginals for \textbf{a} NF-$\kappa$B in WT cells, \textbf{b} NF-$\kappa$B in A20$^{-/-}$ cells, \textbf{c} ATF-2 in WT cells, and \textbf{d} ATF-2 in A20$^{-/-}$ cells.
\textbf{e} CDF of input distribution. \textbf{f} CDF output distribution. \textbf{g} Quantile function of input distribution. \textbf{h} Quantile function of output distribution. The distribution profiles at 30-minute time point have not been shown explicitly to avoid redundant panels. 
In panels \textbf{a}-\textbf{d}, the different colors identify the 13 TNF doses listed in the key (rightmost). In panels \textbf{e}-\textbf{h}, colors and symbols denote the pathway and cell type as labeled. 
}
\label{fs2}
\end{figure}

\section{TNF signaling data analysis}

We adopt the single-cell measurements of the NF-$\kappa$B and ATF-2 signaling pathways stimulated by TNF (Figs.~\ref{fs1}a), reported in Cheong \textit{et al.} \cite{Cheong2011}. We have extracted the reported dose-response statistics for WT and A20$^{-/-}$ cells at 30 minutes and 4 hours after TNF stimulation, and we reproduce them in Figs.~\ref{fs1}b,d. Here, we consider the TNF concentration as the input $X$ and either NF-$\kappa$B or ATF-2 as the output $Z$, depending on the pathway analyzed. For each TNF concentration, denoted $x_i$ (ng/mL), the dataset provides the corresponding conditional mean concentration $\mu_{Z|X}(x_i)$ (a.u.), and conditional standard deviation (SD) $\sigma_{Z|X}(x_i)$ (a.u.) for the outputs in both cell types (see Figs.~\ref{fs1}b,d). We note that these statistics are used to generate Figs.~6c,e in the main text. For numerical evaluation, at each input concentration $x_i$, we approximate the conditional distributions of NF-$\kappa$B and ATF-2 by a Gaussian distribution, defined as,
\begin{eqnarray}
    P(z|x_i) = \mathcal{N}\!\left( \mu_{Z|X}(x_i), \sigma^2_{Z|X}(x_i) \right).
    \label{eq24}
\end{eqnarray}

\noindent The distributions are discretized on a grid bounded by $z \in [z_{\min}, z_{\max}]$, where,
\begin{eqnarray*}
    z_{\min} &=& \max \left( 0, \min_i\left[\mu_{Z|X}(x_i)-4\sigma_{Z|X}(x_i)\right] \right), \\
    z_{\max} &=& \max_i\left[\mu_{Z|X}(x_i)+4\sigma_{Z|X}(x_i)\right],
\end{eqnarray*}

\noindent with normalization $\sum_kP(z_k|x_i)\Delta z=1$. This choice of discretization is used to ensure consistency, where a single common range for the output variable $z$ could be applied to all input concentrations. Moreover, the bounds are set using $\pm 4\sigma_{Z|X}(x_i)$ because, for approximately Gaussian distributions, we assume that the probability mass outside this range is negligible. This ensures that nearly all biologically relevant variation is retained without allocating computational effort to regions where the density is effectively zero. Now, the grid resolution (number of discretized bins) is chosen adaptively according to,
\begin{eqnarray*}
    n_Z = \bigg\lceil \frac{\rho_Z (z_{\max}-z_{\min})}{\min_i[\sigma_{Z|X}(x_i)]} \bigg\rceil,
\end{eqnarray*}

\noindent where, $\rho_Z$ is a tunable discretization density parameter. Scaling $\rho_Z$ with the global output range and the minimum conditional SD ensures that even the narrowest distributions are sampled with sufficient resolution. The ceiling function $\lceil \cdot \rceil$ rounds up the real value to the nearest integer. The stability of this discretization is verified by systematically varying $\rho_Z$ and recalculating all quantities of interest. We find that at $\rho_Z=200$ at 30-minute time point and $\rho_Z=300$ at 4-hour time point, both MI and the 2-WD maintain approximately stable plateaus (Fig.~\ref{fs1}c,e), which confirms numerical robustness.

Given a candidate input distribution $P_X(x)=\{P_X(x_i)\}$, the output marginal distribution is given by $P_Z(z)=\sum_i P_X(x_i)P(z|x_i)$ and the discrete form of MI with bin width $\Delta z = (z_{\max}-z_{\min})/n_Z$ gives,
\begin{eqnarray}
    I(X;Z) = \sum_{i,k} P_X(x_i)P(z_k|x_i) \log_2 \left[ \frac{P(z_k|x_i)}{P_Z(z_k)} \right]  \Delta z.
    \label{eq25}
\end{eqnarray}

\noindent We determine $P_X(x)$ by minimizing $[I(X;Z)-I_{\text{target}}(X;Z)]^2$ subject to $\sum_iP_X(x_i)=1$ and $0 \le P_X(x_i) \le 1$, where $I_{\text{target}}(X;Z)$ denotes the reported values of MI in Cheong \textit{et al.} \cite{Cheong2011}. After optimization, our estimated $I(X;Z)$ matches the reported channel capacity $I_{\text{target}}(X;Z)$ within numerical tolerance. The resulting $P_X(x)$ and $P_Z(z)$ are then used to estimate the 2-WD. The Gaussian approximated $P(z|x)$, the optimal $P_X(x)$ and $P_Z(z)$ at 4-hour time point are given in Figs.~\ref{fs2}a-d for different signaling pathways and experimental conditions. We note that the distribution profiles at 30-minute time point are not shown to avoid redundant panels; their quantitative effects are already summarized through the mean/CV profiles and the dual-fidelity values.

\begin{figure}[!t]
\centering
\includegraphics[width=0.76\columnwidth,angle=0]{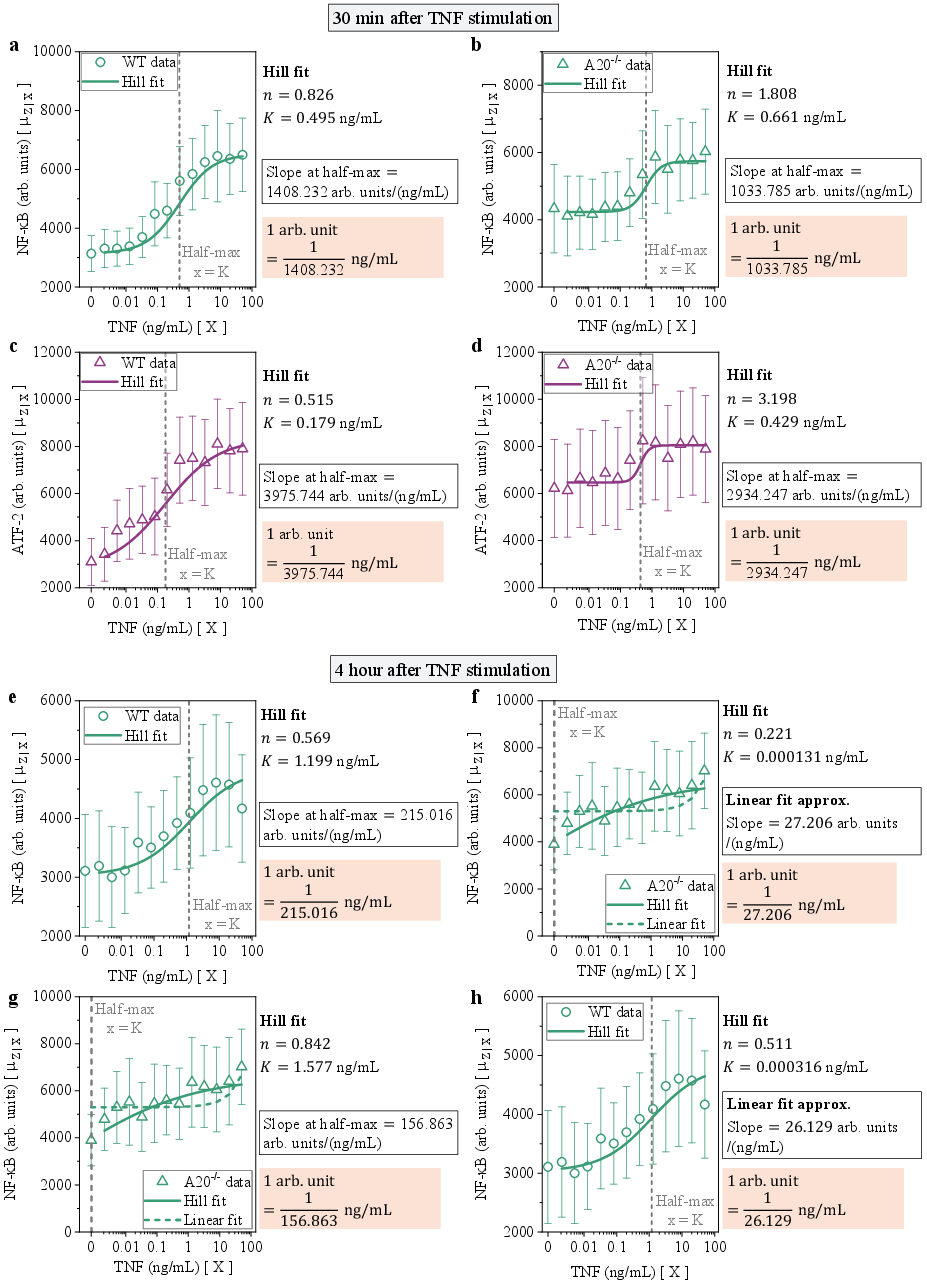}
\caption{\textbf{Dose-response fitting and mapping factors.}
\textbf{a} WT cells with Hill fit and slope at half-maximum for NF-$\kappa$B at 30-minute time point.
\textbf{b} A20$^{-/-}$ cells with Hill fit and slope at half-maximum for NF-$\kappa$B at 30-minute time point.
\textbf{c} WT cells with Hill fit and slope at half-maximum for ATF-2 at 30-minute time point.
\textbf{d} A20$^{-/-}$ cells with Hill fit and slope at half-maximum for ATF-2 at 30-minute time point.
\textbf{e} WT cells with Hill fit and slope at half-maximum for NF-$\kappa$B at 4-hour time point.
\textbf{f} A20$^{-/-}$ cells with linear fit due to the lack of a sigmoidal profile for NF-$\kappa$B at 4-hour time point.
\textbf{g} WT cells with Hill fit and slope at half-maximum for ATF-2 at 4-hour time point.
\textbf{h} A20$^{-/-}$ cells with linear fit due to the lack of a sigmoidal profile for ATF-2 at 4-hour time point.
Slopes define the mapping from arb. units to ng/mL (orange colored block). The mean (symbols) and corresponding error bars are obtained from the mean and standard deviation, respectively, given in Fig.~\ref{f1}b,d. The vertical grey dashed lines indicate $x=K$, the half-maximal dose to the Hill fit. Each panel displays dose-response curves for all 13 TNF doses. 
}
\label{fs3}
\end{figure}

Since both the input and the output are one-dimensional, we use the quantile-based definition of the 2-WD, which is given by \cite{Peyre2019},
\begin{eqnarray}
    W(X,Z) = \sqrt{ \int_0^1 \left( F_X^{-1}(q) - F_Z^{-1}(q) \right)^2 dq},
    \label{eq26}
\end{eqnarray}

\noindent where $F_X^{-1}(q)$ and $F_Z^{-1}(q)$ are the quantile functions of $P_X(x)$ and $P_Z(z)$. Numerically, quantiles are computed from the cumulative distribution functions (CDFs) of $P_X(x)$ and $P_Z(z)$ over 1000 quantile points. The CDFs and quantiles at 4-hour time point are shown in Figs.~\ref{fs2}e-h as representative examples of the procedure used to compute the 2-WD.

Because the input quantiles are in ng/mL while the output quantiles are in a.u., the two must be expressed on the same scale before evaluating the 2-WD using Eq.~(\ref{eq26}). This is necessary because the 2-WD depends explicitly on distances measured in the units of the underlying variable. Our goal here is not to biochemically calibrate fluorescence units, but to ensure that small changes in input and output are compared consistently in the dose range where the response is most sensitive. We therefore define a mapping that aligns output fluctuations with input fluctuations around the point of maximal sensitivity of the dose–response curve. To compute the mapping factor, we fit the experimentally reported discrete dose-response data $\{ \mu_{Z|X}(x_i) \}$ using a continuous Hill function of the form,
\begin{eqnarray}
    R(x) = R_{\min} + \frac{R_{\max}-R_{\min}}{1+ (K/x)^n},
    \label{eq27}
\end{eqnarray}

\noindent where, $x$ is treated as a continuous approximation of the discrete TNF concentrations $x_i$. The fitted function satisfies $R(x_i) \approx \mu_{Z|X}(x_i)$ for all measured points. In this expression, $R_{\max}$ and $R_{\min}$ represent the basal and saturating output concentrations, $K$ is the half-maximal TNF concentration, and $n$ is the Hill coefficient. To obtain a consistent mapping factor, we compute the slope of the fitted curve at the half-maximum point, i.e., $x=K$,
\begin{eqnarray*}
    \frac{dR(x)}{dx} \Bigg|_{x=K} = \frac{n(R_{\max}-R_{\min})}{4K},
\end{eqnarray*}

\noindent This slope provides a standardized local gain of the dose-response curve at half-maximum. The half-maximum point corresponds to the region of highest sensitivity and minimal saturation, which contributes most strongly to both mutual information and geometric comparisons in sigmoidal signaling responses. The reciprocal of this slope, with units of ng/mL per a.u., is therefore used as the mapping factor to convert output quantiles into the same units as the input. Anchoring the conversion to this sensitive region avoids distortions arising from shallow, low-dose responses or saturated, high-dose regimes.

We find that the dose-response curves of both cell types at 30-minute time point and of WT cells at 4-hour time point are well described by the Hill function with an identifiable half-maximal concentration $K$ (Figs.~\ref{fs3}a-d,e,g). In contrast, the dose-response curves of A20$^{-/-}$ at 4-hour time point do not display a clear sigmoidal regime over the measured dose range, leading to poorly constrained Hill fits and spuriously small estimates of $K$ (Figs.~\ref{fs3}f,h). To obtain a robust scale conversion in this case, we instead use a linear approximation $R(x)=a+bx$, compute the slope $b$, and take its reciprocal $1/b$ as the mapping factor (Figs.~\ref{fs3}f,h). After mapping the output quantiles onto the same scale as the input, we measure the 2-WD for NF-$\kappa$B and ATF-2 in both cell types at both time points. The error bars in both MI and the 2-WD are estimated using bootstrap resampling.

We note that alternative strategies exist to place input and output distributions on a common scale before computing the 2-WD, such as rescaling each marginal by its standard deviation or by an empirical dynamic range. However, these normalizations can be dominated by low-gain regions of the dose–response, including low-dose noise and high-dose saturation, where changes in input produce little change in output. This issue is particularly relevant for A20$^{-/-}$ cells at 4-hour time point, which do not exhibit apparent sigmoidal behavior of the dose-response curves over the probed concentration range (Figs.~\ref{fs3}f,h). Our slope-based mapping, therefore, defines a common, gain-based scale for comparing geometric fidelity across conditions.

\begin{figure}[!t]
\centering
\includegraphics[width=1.0\columnwidth,angle=0]{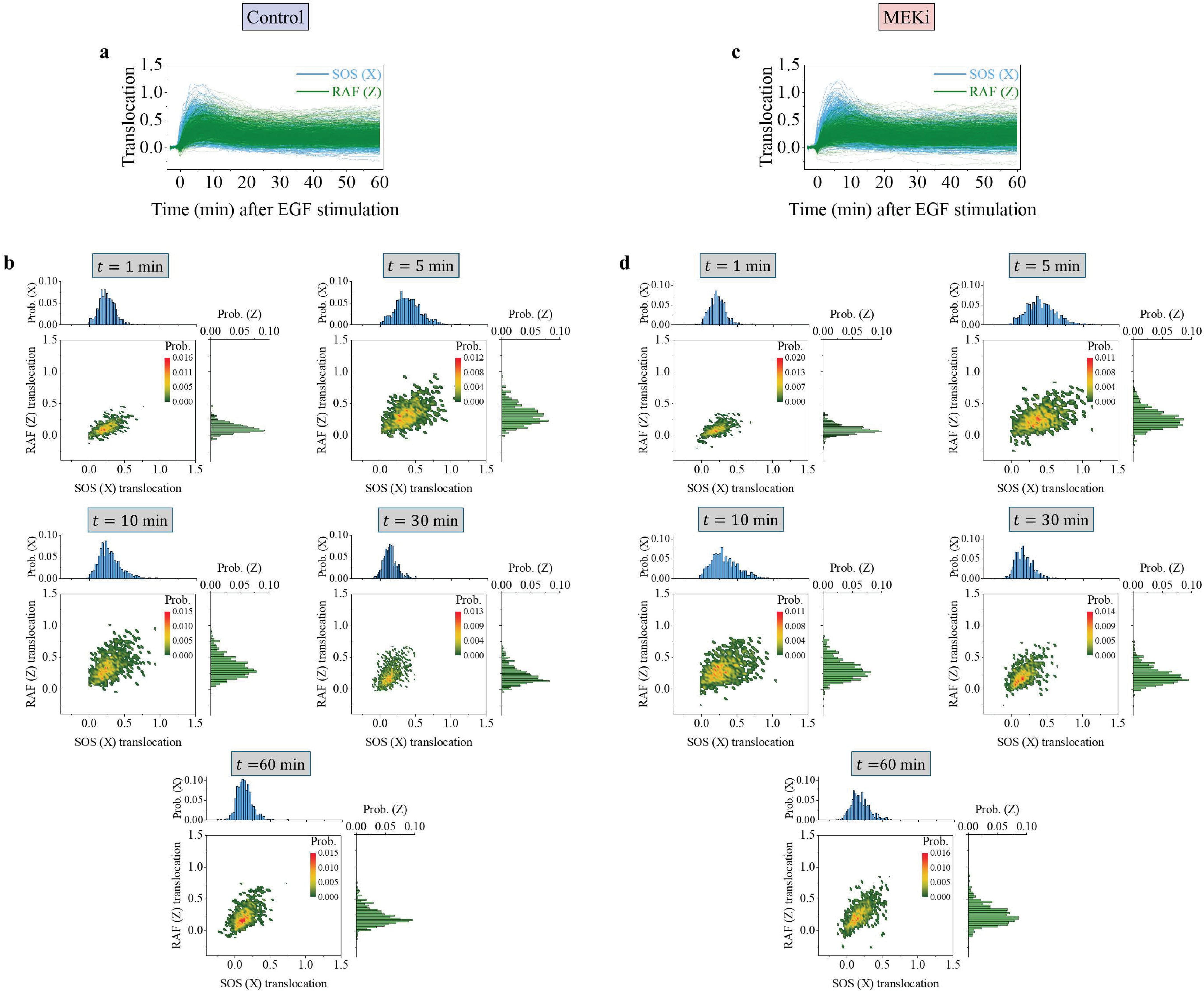}
\caption{
\textbf{Single-cell time-series and the distribution profiles of SOS and RAF translocation.}
\textbf{a} Time-series profiles of SOS and RAF translocation under the control condition.
\textbf{b} Representative marginal and joint distributions of SOS and RAF translocation under the control condition at selected time points.
\textbf{c} Time-series profiles of SOS and RAF translocation under the MEKi condition.
\textbf{d} Representative marginal and joint distributions of SOS and RAF translocation under the MEKi condition at selected time points.
The time-series data are reported by Umeki et al. \cite{Umeki2025}. The control condition contains $n=744$ single-cell trajectories, and the MEKi condition contains $n=706$ single-cell trajectories at each time point. Successive time points represent repeated measurements of the same cells. The marginal and joint distribution profiles are generated from the corresponding single-cell time-series data.
}
\label{fs4}
\end{figure}

\clearpage

\end{bibunit}

\end{document}